\documentclass[%
 reprint,
 superscriptaddress,
amsmath,amssymb,
aps,
pra,
longbibliography,
noeprint
]{revtex4-2}

\usepackage{graphicx}
\usepackage{dcolumn}
\usepackage{bm}
\usepackage[hidelinks]{hyperref}
\usepackage{overpic}

\usepackage{lineno} 
\usepackage[caption=false]{subfig}
\usepackage{soul} 
\usepackage{physics}
\newcommand{\mum}{\mu\textrm{m}}

\usepackage{enumitem,amssymb}
\newlist{todolist}{itemize}{2}
\setlist[todolist]{label=$\square$}
\usepackage{pifont}

\newcommand{\refeq}[1]{(\ref{#1})}
\newcommand{\refig}[1]{Fig.~\ref{#1}}
\newcommand{\intd}[1]{\int\!d{#1}\,}
\def\mc{\mathcal}
\def\mr{\mathrm}

\def\fr{\frac}

\graphicspath{{Figures/}} 

\usepackage{xstring}
\usepackage{catchfile}
\usepackage{xcolor}

\newcommand{\affA}{%
\affiliation{Experimental Quantum Information Physics Unit, Okinawa Institute of Science and Technology, Onna, Okinawa, 904-0495, Japan}
     }
\newcommand{\affB}{%
\affiliation{Quantum Material Science Unit, Okinawa Institute of Science and Technology, Onna, Okinawa, 904-0495, Japan}
	}

\begin{document}

\title{How to integrate a miniature optical cavity in a linear ion trap: shielding dielectrics and trap symmetry}

\author{Ezra Kassa$^\dagger$}
\email{ezra.kassa@oist.jp} 
\author{Shaobo Gao}
\thanks{These two authors contributed equally.}
\author{Soon Teh}
\affA

\author{Dyon van Dinter}
\affB

\author{Hiroki Takahashi}
\email{hiroki.takahashi@oist.jp}
\affA


\begin{abstract}
One method of scaling up quantum systems is to adopt a modular approach. In the ion trap architecture, an efficient photonic interface between independent linear ion traps would allow for such expansion. To this end, an optical cavity with a small mode volume can be utilised to enhance the photon emission probability from the ion. Miniature fibre-based Fabry-Perot cavities have been integrated into three-dimensional Paul traps that hold a single ion, whereas an efficient interface between an optical cavity and a linear trap that can keep multiple ions has remained elusive. This presents a barrier for combining the benefits of the motional coupling in a chain of ions with optical interface between ion traps.
In this paper, we show that simple electrically conductive shielding of the fibres could provide substantial advantage in mitigating the adverse effects of stray charges and motional heating by dielectrics. We also reveal that the conductive shields are not compatible with the conventional radio frequency (rf) drive in ion traps but using two rf signals with opposite phases can solve this issue. Furthermore the role played by the symmetry of the electrodes when incorporating an element that disrupts the translational symmetry of a linear trap is elucidated analytically. As a result it is realized that two-dimensional implementation of a linear ion trap such as a surface trap is inherently not suitable for integrating a shielded miniature optical cavity due to the lack of geometrical symmetry. Based on the insights obtained through the analysis, we identify essential components and a design strategy that should be incorporated in a linear ion trap for successful integration of a miniature optical cavity. 
\end{abstract}

\maketitle

\section{Introduction}

Laser-cooled atomic ions trapped in radio-frequency (rf) Paul traps are ideal stationary qubits with their long trapping lifetimes, long coherence times \cite{Harty:14} and high single- and multi-qubit gate fidelities \cite{Harty:14,Gaebler2016}. So far, however, they have been lacking an efficient interface with flying photonic qubits that would enable scaling up ion-based quantum processors and long-distance quantum communication via photonic interconnects \cite{Leibdried:03}. Cavity quantum electrodynamics (cQED) could provide a solution for this outstanding problem through enhanced photon emission of the ions coupled to an optical cavity. Along this line the integration of optical cavities in ion traps have been experimentally pursued in the last three decades. 
In recent years, the integration of miniature optical cavities in the needle- and endcap-type ion-traps, both of which can only stably trap a single ion at a time, have found success \cite{Steiner:13, Takahashi:14}. Nonetheless the challenge of combining the advantages of mutually coupled ions in a linear Paul trap with cQED in the strong coupling regime remain unsolved \cite{Bransdstatter:13, Ong2020, Teller2023}.

In this paper, we examine the issues that originate from the integration of a miniature fibre-based Fabry-Perot cavity (FFPC) with a linear Paul ion trap through numerical and analytical studies. We investigate adversary effects caused by the FFPC such as stray electric charges, disturbance to the trapping potential, and increased heating rates of the ion's motion. In particular we are interested in the effects of electrically conductive shielding around the fibres. We have found that the shielding mitigates some of the adversary effects but introduces a new disturbance to the trapping potential. The latter can be circumvented by changing the way in which the rf signal is applied to the trap. We discuss the cause of this disturbance and the solution thereof in light of the geometrical symmetry of the ion trap.

The paper is organised as follows: In Section~\ref{review}, previous studies regarding the integration of optical cavities in ion traps are reviewed. In Section~\ref{sec:cavity_geom}, consideration on the geometry of the optical cavity is given in the perspective of cQED, particularly for achieving a large cooperativity parameter. In Section~\ref{sec:model_methods}, the model of the ion trap and the methods used in the numerical simulations in the following sections are introduced. In Section~\ref{sec:mitigating_effects}, the effects of conductive shields around the FFPC that mitigate the adverse effects of stray charges and motional heating are numerical investigated. In Section~\ref{sec:price}, the deformation of the trapping potential due to the introduction of the conductive shields is studied. Two different driving schemes for the rf signal are employed in the simulation to reveal their difference in the cancellation of the rf field on the trap axis. This aspect is analytically elucidated in Section~\ref{sec:trap-symmetry} in light of the geometrical symmetry of the trap with additional results for planar surface traps. Section~\ref{sec:discussion} concludes the paper with discussion of the obtained results and proposes a strategy for achieving successful integration of a miniature optical cavity in a linear ion trap.     


\section{Review of previous studies}
\label{review}

The first series of experiments that successfully coupled ions with an optical cavity used macroscopic mirrors with a cavity length large enough that the mirrors did not significantly hamper the trapping capabilities. These experiments reported groundbreaking results such as the demonstration of a single ion as a nanoscopic probe of an optical field \cite{Guthohrlein2001a} and the generation of single photons on demand \cite{Keller:04}.
As the coupling strength, $g$, scales with the cavity mode volume as $g \propto 1/\sqrt{V}$, the large cavity length of the aforementioned experiments meant that the cavity mode volumes were also large. 
Consequently, the coupling strengths were exceeded by the decoherence rates: these set-ups could not be used as efficient quantum interfaces to link up multiple ion-trap nodes, and the pursuit for ions strongly coupled to an optical cavity continued. 

Other experiments took a different approach: increase the cavity length close to the concentric regime to minimise the mode volume at the expense of susceptibility to mechanical instability. This led to the demonstration of a tunable entanglement between an ion and a photon \cite{Stute:12} where a 2~cm long cavity mode's waist has been reduced to 13 $\mum$, but remained far from the  strong coupling regime with $(g,\gamma) = 2\pi(1.4, 11)$ MHz, with $\gamma$ representing the atomic spontaneous decay rate.

Around the same time, optical cavity mirrors on the facets of fibre tips were developed \cite{Hunger:10}, significantly reducing the size of integrable cavities with neutral atom traps whilst retaining good optical access. These were adapted for use with ion traps \cite{Takahashi:14} allowing the customisation of the cavity lengths and the radii of curvature of the cavity mirrors, leading to a reduction in mode volume by two orders of magnitude.
The use of these fibre-based Fabry-Perot cavities (FFPCs) found success in a number of  experiments \cite{Steiner:13, Meyer2015, Ballance2017, Takahashi2017, Kassa2018} leading up to the strong coupling of a single ion to an optical cavity \cite{Takahashi:20}.
In these experiments using FFPCs, the endcap-type\cite{Takahashi2017, Kassa2018, Takahashi:20} and the needle-type\cite{Steiner:13, Meyer2015, Ballance2017} ion traps were used. By design both types have a single rf-null point in their electric potentials. Hence they are only suitable for trapping a single ion without excess micromotion.

In order to maintain the benefits of ions as qubits whilst scaling-up by creating a photonic link between nodes hosting these ions, each ion trap node must be able to store multiple ions.
In this sense, the aforementioned experiments that used ion traps with single rf-null points were of limited practical use, and the integration of miniature cavities in linear ion traps, which can store multiple ions in a string and perform local multi-qubit logic, was highly desired. This would allow large scale entanglement distribution and thereby solve the scalability limitations that has been hindering advances in this field of research and technology.
For this reason, the integration of an FFPC in a linear ion trap has been also pursued in the last decade \cite{Bransdstatter:13,Pfister2019,Ong2020,Schuppert2020}. However, none of these experiments has so far achieved stable strong coupling between the ion and cavity. In \cite{Ong2020}, the authors investigated the effects of charging on the dielectric surfaces of the fibres and found that the ion can be used as an effective probe for these surface charges. On the other hand, it seems that this same surface charging effect ultimately hindered them from coupling the ions stably to the cavity \cite{Schuppert2020}. It is worth noting that the same experimental group also found that the dielectric mirror surfaces situated closely to the trapped ion can cause significant motional heating, orders of magnitude more so than with metal surfaces \cite{Teller2021}. Even though this is currently not a critical obstacle for coupling an ion to a cavity in a linear trap, it will be highly problematic once a miniature cavity is integrated in a linear trap and motion-based quantum gates are performed between ions in the proximity of the cavity mirrors.
Similarly, \cite{Pfister2019} reports that ions in the linear trap could not be reliably translated to be coupled to an FFPC and find that charge buildup on the FFPC creates a potential barrier repelling the ion from the cavity region.  

Comparing the attempts to integrate cavities in the single rf-null traps \cite{Steiner:13,Takahashi2017} and those in linear traps \cite{Bransdstatter:13, Pfister2019, Ong2020}, apart from the obvious difference in the trap geometries one can point out a difference in the degrees of electric shielding employed for the fibres. In the former the fibres were enclosed in metallic cylinders and only the front facets were exposed. In the latter, the fibres were placed without any shielding and all the surfaces of the fibres were exposed.
One can argue that this may explain the difference in the severity of the charging effects experienced in the two approaches.

Even though charging and discharging dynamics on dielectric surfaces in ion traps have been hardly understood due to their high complexities, in this paper we adopt a design that encloses the fibres in electrically conductive shields as a potential remedy for the issue of stray charges in linear traps. We numerically examine how this conductive shielding of the fibres affects the trapping potential and how much it mitigates the effects of deposited charges and heating from the mirror surfaces.    

Lastly let us mention ref.~\cite{Podoliak2016} where numerical investigation similar to the current work concerning integration of FFPCs in ion traps was conducted. In \cite{Podoliak2016}, various different trap geometries were considered, including the blade and surface linear traps as in the current work. However, in \cite{Podoliak2016} conductive shielding for the fibres was not introduced in the linear traps. As is shown in Section~\ref{sec:mitigating_effects}, without conductive shields, dielectric fibres alone would not hamper the stable trapping of ions as far as simulation is concerned. The experimental studies, however, have revealed that unshielded fibres could be a source of critical instability in the ion traps as discussed above. In the current work, we take the conductive shielding of the fibres to be an essential design choice and it constitutes the main difference from \cite{Podoliak2016}. Besides we numerically evaluate the motional heating of the ion in the presence of the fibre dielectrics, which was also missing in \cite{Podoliak2016}.

\section{Consideration on the cavity geometry}
\label{sec:cavity_geom}

\begin{figure}[!h]
    \centering
    \includegraphics[width=1\linewidth]{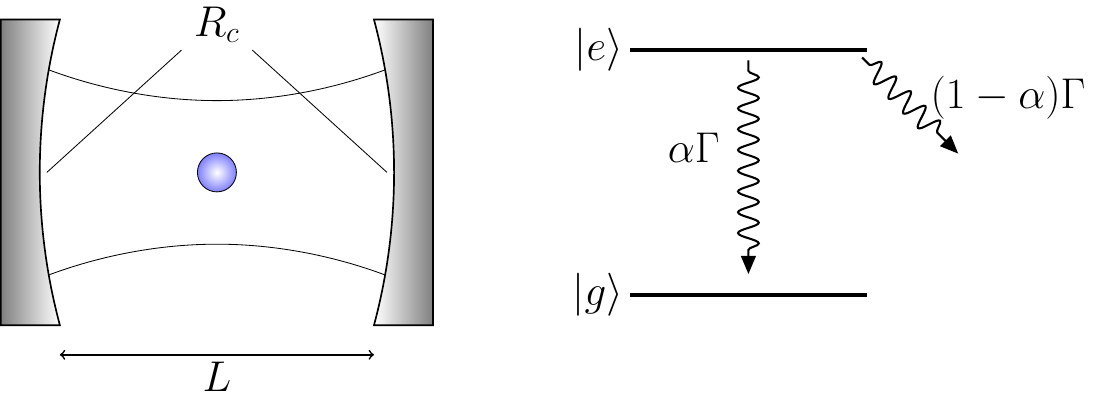}
    \caption{Left: cQED model with an ion trapped in an open Fabry-Perot cavity. Right: level scheme of the ion. The excited state $\ket{e}$ has a radiative population decay rate of $\Gamma$. The decay channel $\ket{e}\rightarrow\ket{g}$ has a branching ratio of $\alpha$. $(1-\alpha)\Gamma$ gives the rate of the decay to the states other than $\ket{g}$.}
    \label{fig:cQED}
\end{figure}
We begin by motivating the need for a miniature size of  the optical cavity from the view point of cQED. This in turn necessitates a miniature size of the ion trap. Optical cavities are inherently incompatible with ion traps. The fluctuating electric potential on these dielectric materials hampers stable trapping of the ions \cite{Harlander2010, Brownnutt2015, Teller2021}. Therefore, one would ideally position the dielectric materials that constitute the cavity far from the trapping region. This almost precludes the use of any type of optical cavity that uses an evanescent field to couple the ion \cite{Armani2003,Ruddell2020}. Hence in this paper we focus on the Fabry-Perot type optical cavities consisted of two separate mirrors with empty space between them. In this case the large distance between the ion and dielectrics is equivalent to a large cavity length. 

On the other hand, efficient ion-cavity coupling requires a different figure of merit. Let us consider a model depicted in \refig{fig:cQED}. A single ion with energy eigenstates $\ket{e}$ and $\ket{g}$ is coupled to a Fabry-Perot cavity formed by two identical concave mirrors of a radius of curvature $R_c$ and with a cavity length $L$ . The cavity is resonant with the transition frequency between $\ket{e}$ and $\ket{g}$. The excited state $\ket{e}$ has a radiative population decay rate of $\Gamma$ while it is assumed that the branching ratio of the $\ket{e} \rightarrow \ket{g}$ channel is $\alpha$ so that the decay rate of this channel is given by $\alpha\Gamma$ (see \refig{fig:cQED}). As is well known, most coherent processes between an ion and a cavity are characterized by the cooperativity parameter \cite{Goto:19}:
\begin{align}
    C = \frac{g^2}{\Gamma\kappa},
\end{align}   
where $g$ is the coherent ion-cavity coupling rate, $\kappa$ is the amplitude decay rate of the cavity. 
The probability that the ion emits a photon into the cavity mode once it is promoted to $\ket{e}$ is given by the following formula \cite{Goto:19}.
\begin{align}
    P_{emi} = \frac{2C}{2C+1}.
\end{align}
In order for this emission to be predominant over the emission into the free space, here we set $C \gtrsim 10$ as a guideline condition that should be satisfied for efficient ion-cavity interface.

One would design the cavity to maximize the cooperativiy parameter within the limits of practical constraints. It can be shown that $C$ can be expressed in the cavity finesse $\mathcal{F}$, the cavity waist size $\omega_0$, the resonance wavelength $\lambda$ and the branching ratio $\alpha$ as follows:
\begin{align}
    C = \frac{3\alpha\mathcal{F}}{\pi^3}\qty(\frac{\lambda}{\omega_0})^2. \label{cooperativity}
\end{align}
Furthermore the waist size of the fundamental TEM00 mode of the cavity is given by
\begin{align}
    \omega_0 = \qty(\frac{\lambda}{\pi})^{1/2}\qty(\frac{L}{2}\qty(R_c-\frac{L}{2}))^{1/4}. \label{caivty-waist-size}
\end{align}
Substituting \refeq{caivty-waist-size} to \refeq{cooperativity}, one obtains
\begin{align}
    C &= \frac{3\mathcal{F}}{\pi^2}\frac{\alpha\lambda}{\sqrt{\frac{L}{2}\qty(R_c-\frac{L}{2})}}
    = \frac{3\mathcal{F}}{\pi^2}\frac{\eta}{R_c\sqrt{\xi(1-\xi)}}, \label{cooperativity2}
\end{align}
where $\eta = \alpha\lambda$ and $\xi = L/(2R_c)$ with the latter giving the cavity length normalized by its concentric limit $2R_c$. Note that $\eta$ has the dimension of length and is solely a function of the atomic properties of the ion. The values of $\eta$ and other relevant parameters for different ionic species and transitions are listed in Table \ref{tab:ion_params}. In contrast to $\eta$, the rest of the parameters in \refeq{cooperativity2} depends only on the cavity characteristics. The cavity finesse $\mc{F}$ is determined by the reflectivies of the mirrors and optical losses in the cavity, and hence basically free from the geometry of the cavity \footnote{The cavity finesse could indirectly depend on the cavity geometry when the clipping losses of the cavity field on the mirrors become non-negligible.}. The geometry of the cavity is reflected through $R_c$ and $\xi$ in \refeq{cooperativity2}.

\begin{table*}[tb]
    \centering
    \begin{tabular}{|c|c|c|c|c|} 
     \hline
     Species & Transition & Wavelength $\lambda$ ($\mum$) & Branching ratio $\alpha$  & $\eta = \alpha\lambda$ ($\mum$)\\
     \hline
     $\mr{Ca}^+$ & $S_{1/2}-P_{1/2}$ & 0.397 & 0.936 & 0.372 \\ 
     \hline
     $\mr{Ca}^+$ & $D_{5/2}-P_{3/2}$ & 0.854 & 0.059 & 0.050 \\ 
     \hline
     $\mr{Sr}^+$ & $S_{1/2}-P_{1/2}$ & 0.422 & 0.941 & 0.397 \\ 
     \hline
     $\mr{Sr}^+$ & $D_{5/2}-P_{3/2}$ & 1.033 & 0.053 & 0.055 \\
     \hline
     $\mr{Ba}^+$ & $S_{1/2}-P_{1/2}$ & 0.493 & 0.729 & 0.359 \\ 
     \hline
     $\mr{Ba}^+$ & $D_{5/2}-P_{3/2}$ & 0.614 & 0.215 & 0.132 \\
     \hline
     $\mr{Yb}^+$ & $S_{1/2}-P_{1/2}$ & 0.370 & 0.995 & 0.368 \\
     \hline
     $\mr{Yb}^+$ & $D_{3/2}-D[3/2]_{1/2}$ & 0.935 & 0.018 & 0.017 \\
     \hline
    \end{tabular}
    \caption{A list of wavelengths, branching ratios and values of $\eta$ for relevant transitions in major ionic species.}
    \label{tab:ion_params}
\end{table*}

\begin{figure}[tb]
    \centering
    \subfloat[]{\includegraphics[width=0.8\linewidth]{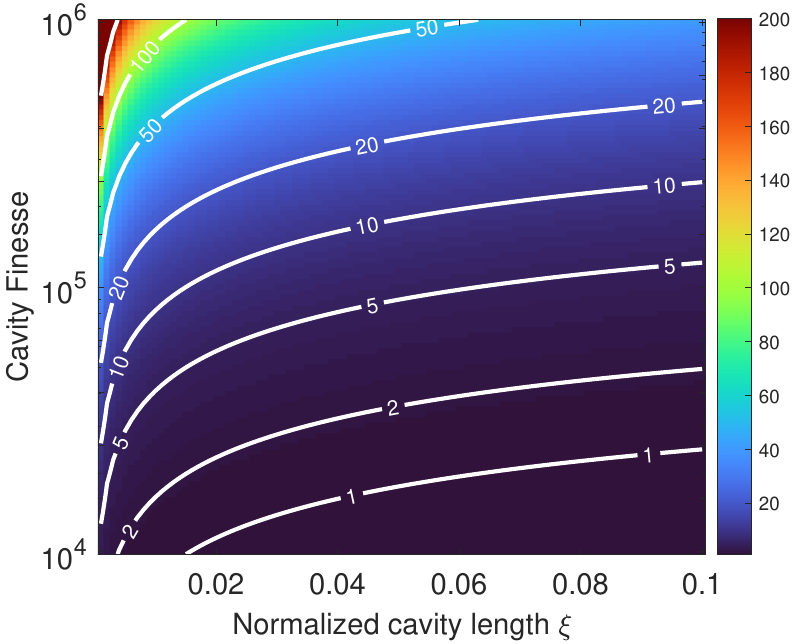}\label{fig:C_eta_SP}} \\
    \subfloat[]{\includegraphics[width=0.8\linewidth]{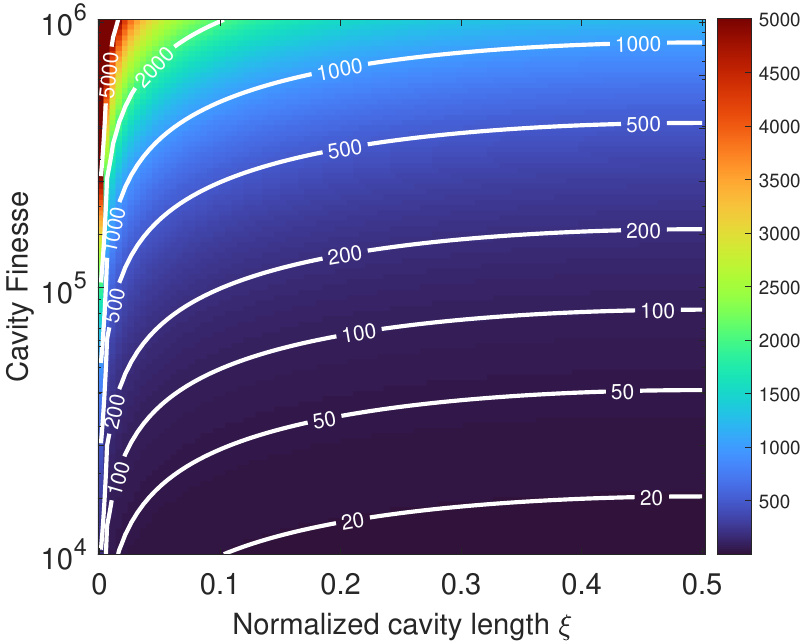}\label{fig:C_eta_MF}}
    \caption{2D plots of $C/\eta$ as a function of the cavity finesse and normalized cavity length in the cases of \textbf{(a)} $R_c = 25\,\mr{mm}$ and \textbf{(b)} $R_c = 500\,\mum$ respectively. Note the differences between two plots in the color scaling and the ranges of the x-axes. The white solid lines are the contours at the labeled values.}
    \label{fig:C_eta}
\end{figure}
Now let us estimate the necessary cavity geometry to achieve $C \gtrsim 10$ in practice.
With regard to $R_c$, there are categorically two different regimes depending on the way to fabricate the mirrors. One way is to machine and polish the mirror based on mechanical abrasion of a glass substrate. The highest quality is obtained by a  technique called the superpolishing \cite{Nelson2019}. This method produces excellent surface smoothness but the lowest attainable value for $R_c$ is limited at around $\sim10^4\,\mum$. Another way is to utilize some form of microfabrication either by the $\mr{CO}_2$ laser ablation \cite{Takahashi:14,Ott:16} or chemical etching \cite{Jin2022,Wachter2019}. In this way $R_c$ on the order of $10^2\,\mum$ is possible. The achievable cavity finesse is comparable in both cases \cite{Rempe1992,Jin2022}. In \refig{fig:C_eta_SP} and \ref{fig:C_eta_MF}, $C/\eta$ is plotted for typical values of $R_c$ in the superpolished and microfabricated mirrors respectively.
Even though there is a variation of $\eta$ among different ion species and transitions (see Table~\ref{tab:ion_params}), the parameter space where $C/\eta \gtrsim 100$ is satisfied should be targeted for achieving $C \gtrsim 10$.
It is clear from \refig{fig:C_eta_SP} that with $R_c = 25~\mr{mm}$, this parameter space is highly limited, only achievable with a finesse close to $10^6$ and $\xi < 0.01$ which translates to $L < 500 ~\mum$ for $R_c = 25~\mr{mm}$. Note that one can achieve the same at the other end of the cavity length where $1-\xi < 0.01$ as $C$ is symmetric around $\xi = 0.5$. However, achieving this concentric limit is challenging as the clipping losses of the cavity field at the mirrors become prominent and the cavity becomes more susceptible to mechanical instability \cite{Gao2023}. In contrast, in the case of the microfabricated mirrors with $R_c = 500\,\mum$, the required condition for $C/\eta \gtrsim 100$ is significantly more relaxed as seen in Figure~\ref{fig:C_eta_MF}. $C/\eta \gtrsim 100$ can be achieved at any $\xi$ with a moderate cavity finesse below $10^5$. 

The above example clearly shows that microfabricated mirrors are advantageous for obtaining high cooperativity. Numerically this advantage stems from the small $R_c$ on the order of $10^2~\mum$.
Naturally the cavity length is limited on a similar length scale due to the stability limit ($L < 2R_c$). 
Previous studies demonstrated that a cavity length of $200 - 400~\mum$ is compatible with stable ion trapping in the endcap and needle traps \cite{Steiner:13, Takahashi:20}. Encouraged by these results, in this paper we also assume a cavity length of a few hundred $\mum$ and consider adaptation of such a miniature cavity in the linear trap geometry. 
On a separate note, let us point out that miniature-sized optical cavities are also in demand for the scalability of the ion trap system. In the quantum CCD architecture \cite{Kielpinski2002}, microfabricated electrodes are used in the ion trap at high density. It would be difficult to fit bulky superpolished mirrors in such ion traps. In addition their relatively large dielectric surface areas would make the ion trap more susceptible to stray charging. The small footprints and reduced dielectrics of microfabricated mirrors are preferred in these regards as well.    

\section{The model and methods}
\label{sec:model_methods}

\subsection{Ion trap and cavity geometry}
\label{subsec:ion_trap_geom}
\begin{figure}[h!]
    \centering
    \subfloat[]{\includegraphics[width=\linewidth]{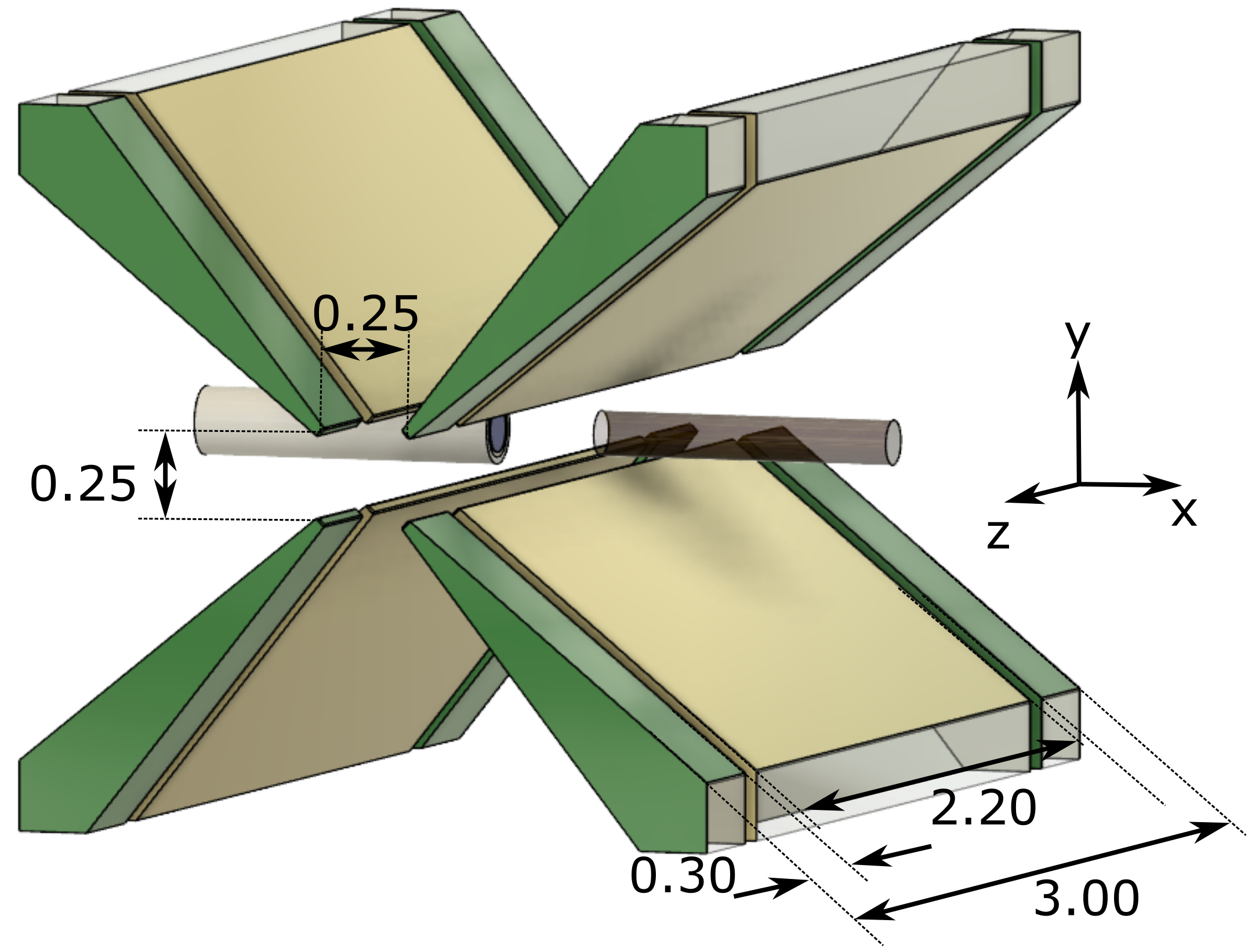}
    \label{fig:3D_model}}
    
    \subfloat[]{\includegraphics[width=\linewidth]{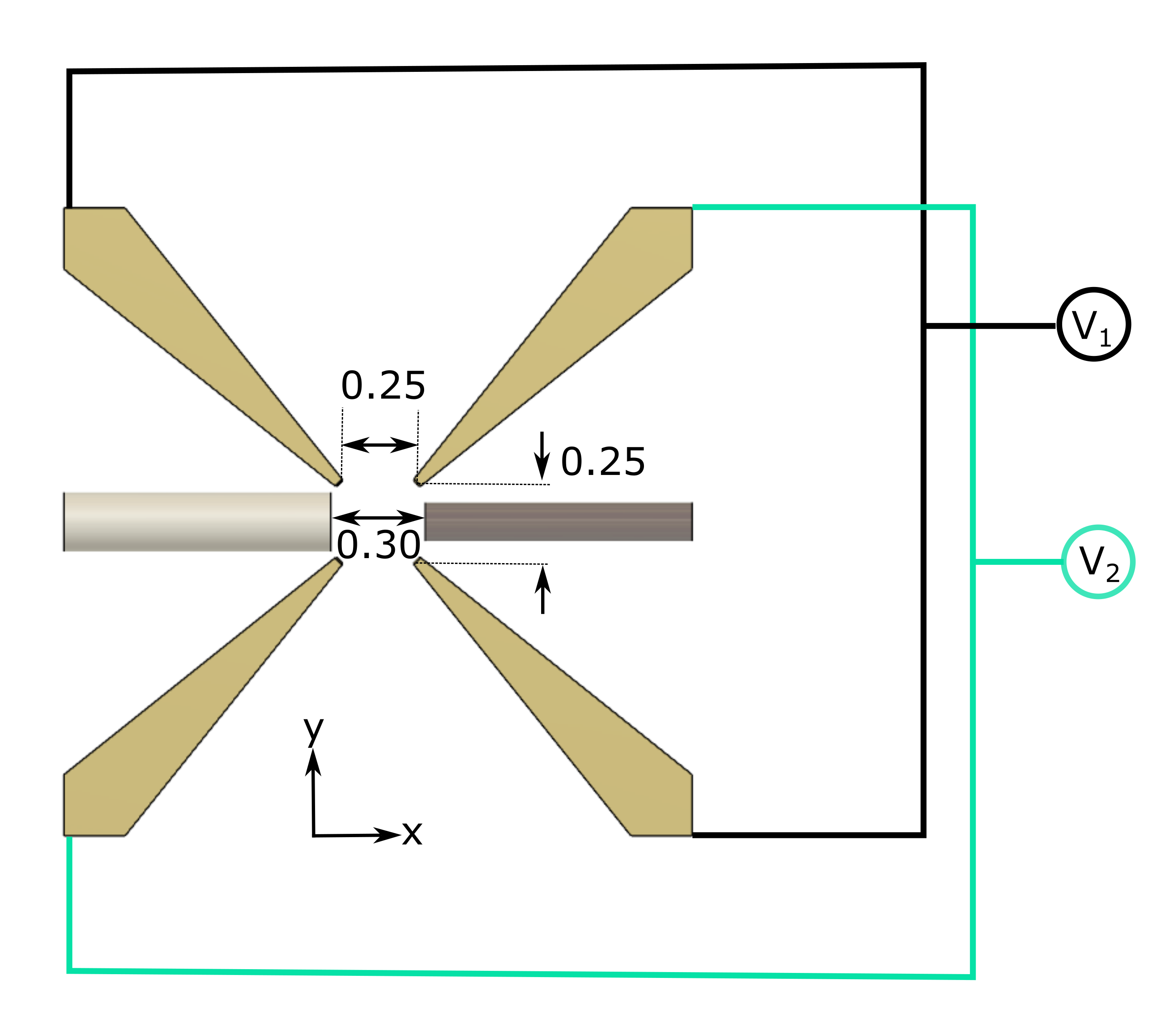}
    \label{fig:trap_xy_view}}
    \caption{\textbf{(a)} The model of a linear ion trap used in the numerical simulations. The central rf electrodes and the endcap dc electrodes are colored yellow and green respectively. For an illustrative purpose we include a conductive shield around the left-hand-side fibre only. \textbf{(b)} Cross sectional view of the trap and FFPC. Rf signals $V_1$ and $V_2$ are applied to the rf electrodes. The tip of electrodes has a width of 22 $\mum$ and their tapering angle is 12.7$^\circ$. All the dimensions are in millimeter. 
    }
    \label{fig:Geom}
\end{figure}

\begin{figure}[tb]
    \centering
        \begin{minipage}{0.45\linewidth}
            \centering
            \subfloat[]{\includegraphics[height=\linewidth]{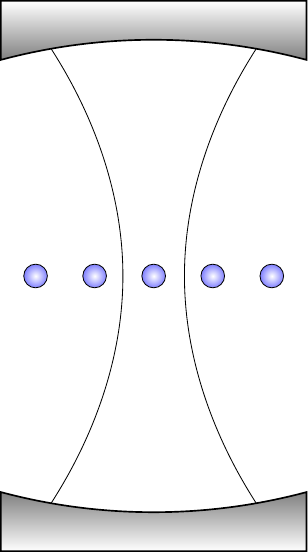}
        \label{fig:cavity_trans}}
        \end{minipage}       
        \hspace{0.5cm}
        \begin{minipage}{0.45\linewidth}
            \centering
        \subfloat[]{\includegraphics[width=\linewidth]{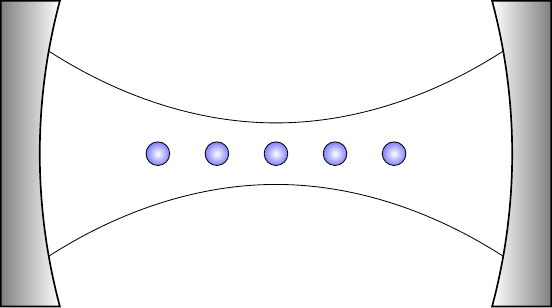}
        \label{fig:cavity_long}}
        \end{minipage}
    \caption{\textbf{(a)} Transverse and \textbf{(b)} longitudinal orientation of the cavity with respect to the ion string.}
    \label{fig:cavity_orient}
\end{figure}

In this paper, we are primarily concerned with a \textit{blade-style} linear ion trap shown in \refig{fig:Geom}. This model is used in the numerical simulations presented in the subsequent sections \ref{sec:mitigating_effects} and \ref{sec:price}.  
The linear trap is modeled as four sets of blades. Each set is comprised with a wide central rf electrode sandwiched by two dc endcap electrodes.
When incorporated in the simulation, the FFPC fibres are modeled as cylinders of silica glass of diameter 125~$\mum$ and are transversely positioned at the trap centre.
The cavity length is set at 300~$\mum$ unless stated otherwise. In some of the simulations, electrically conductive shielding is introduced around the fibres to study its effects. The shield is modeled as a tube of conductor with an inner and outer diameters of 150 and 190~$\mum$ respectively. The cavity axis and the trap axis are along the $x$ and $z$ axes respectively. In the following, we frequently use \textit{axial} and \textit{transverse} directions of the trap which refer to the directions along the $z$ axis and in the $xy$ plane respectively.  

Let us comment on the orientation of the cavity with respect to the ion string. Throughout this paper we only consider the transverse orientation where the cavity is aligned perpendicularly to the ion string as depicted in \refig{fig:cavity_trans}.
Alternatively one could also consider the longitudinal orientation as in \refig{fig:cavity_long}. The longitudinal orientation of the cavity does not violate the axial symmetry of the ion string. In comparison to the transverse configuration where the axial symmetry is violated by the presence of the mirrors, the longitudinal orientation thus has an advantage that it creates less disturbance to the trapping potential \cite{Podoliak2016}. In fact this configuration was experimentally \cite{Begley2016,Teller2023} and numerically \cite{Lee:2019} investigated. The longitudinal orientation, however, has a critical drawback that the trapping region is limited by the cavity length which needs to be a few hundred micrometers as shown in Section \ref{sec:cavity_geom}. Therefore this cavity orientation is not compatible with a large ion trap or complex quantum CCD architecture, and hence not scalable. Besides the mirrors are always within a distance of a few hundred micrometers from the ions. As described in Section \ref{subsec:heating}, the dielectric surfaces of the mirrors cause excessive motional heating to the ions. In the longitudinal cavity configuration this heating cannot be mitigated unless the cavity length is increased or the cavity is put under a cryogenic temperature \cite{Teller2023}. On the ohter hand, in the transverse cavity configuration the ions can be moved away from the mirrors to a position where the heating from the mirrors is negligible, and high-fidelity quantum logics using the motional degrees of freedom can be performed there. In summary, the transverse cavity orientation is more compatible with and favored in quantum information processing using trapped ions, and we believe that realizing it is a necessary step towards achieving scalability in the ion trap system.   

\subsection{Driving schemes for the rf signals}
\label{subsec:rf_drive}

The transverse trapping potential is created by applying rf signals to the rf electrodes. 
The standard way of application is to use a single sinusoidal signal $V_{1} = 2*V_0 \cos{(\Omega_{\textrm{rf}} t)}$ on one of the opposing pairs of the rf electrodes whilst the other pair is held at the rf-ground ($V_2 = 0$) (see \refig{fig:trap_xy_view}).
Since a single rf signal is used, we call this way of application \textit{single-rf drive}. 
Alternatively, we can drive the electrodes with two rf signals 180${}^\circ$ out of phase to each other. That is, $V_{1} = V_0 \cos{(\Omega_{\textrm{rf}} t)}$ and $V_{2} = -V_0 \cos{(\Omega_{\textrm{rf}} t )}$ in \refig{fig:trap_xy_view}. We call this way of application \textit{dual-rf drive}.
The single-rf drive is technically easier to implement and more prevalently used in ion trap experiments than the dual-rf drive. However the latter can be advantageous in a particular trap geometry \cite{Teller2023} or for positioning of the transverse trapping field \cite{Herskind2009b}. In this paper we reveal that the single- and dual-rf drives bring about fundamentally distinct consequences to the trapping potential when they are applied in the presence of the conductive shield for the FFPC. This will be discussed in details in Section~\ref{sec:price} and \ref{sec:trap-symmetry}.

\subsection{Simulation of the trapping potential}
\label{subsec:simulation}

The trapping potential $\Phi$ for the ion is comprised of two contributions:
\begin{align}
    \Phi(x, y, z) = \Phi_{\mr{pseudo}}(x, y, z) + \Phi_{\mr{dc}}(x, y, z). 
    \label{total_pot}
\end{align} 
$\Phi_{\mr{dc}}(x, y, z)$ is produced by the dc voltages on the endcap electrodes and responsible for the axial confinement of the ion. 
On the other hand the transverse confinement is provided by the pseudo-potential $\Phi_{\mr{pseudo}}(x, y, z)$ which is a time-averaged effective potential due to the rf electric field \cite{Leibdried:03}:
\begin{align}
    \vec{E}(x, y, z, t) = \vec{E}_0(x, y, z)\cos\Omega_{\mr{rf}}t. \label{E_rf}
\end{align}
$\Phi_{\mr{pseudo}}(x, y, z)$ is  calculated from the following formula.
\begin{align}
    \Phi_{\mr{pseudo}}(x, y, z) = \frac{e\abs{\vec{E}_0(x, y, z)}^2}{4m\Omega_{\mr{rf}}^2}. \label{pseudo_pot}
\end{align}
Here $e$ is the elemental charge, $m$ is the mass of the ion.
In the simulations, $\Phi_{\mr{pseudo}}(x, y, z)$ and $\Phi_{\mr{dc}}(x, y, z)$ are calculated separately using finite element method software COMSOL Multiphysics\textsuperscript{\textregistered}. As is evident in \refeq{pseudo_pot}, in order to calculate the pseudo-potential, it is not necessary to simulate time dependent electric fields but it is sufficient to calculate a snapshot of the electric field distribution at a particular moment. Therefore for both $\Phi_{\mr{pseudo}}(x, y, z)$ and $\Phi_{\mr{dc}}(x, y, z)$ we run simulations to obtain static electric fields under the corresponding boundary conditions.

Throughout this paper, $V_0 = 30 V$ and $\Omega_{\textrm{rf}} = 2\pi \cdot 20$ MHz are used, and the mass of ${}^{40}\mr{Ca}^+$ is assumed when calculating the pseudo-potential. Secular frequencies of the potentials are extracted by fitting cross sections with parabolas. With the above settings, we obtain radial secular frequencies of $(\omega_x, \omega_y) = 2\pi\,(3.1, 3.1)$~MHz without the FFPC. We will see that these secular frequencies can change in the presence of the FFPC.

\section{Mitigating effects of the conductive shield}
\label{sec:mitigating_effects}
In this section, we study the role played by the conductive shield around the optical fibres. The shield can mitigate some of the adversary effects caused by the presence of the fibres. 

\subsection{Charging effects}
\label{subsec:charging}

\begin{figure*}[t!]
    \centering
    \subfloat[]{\includegraphics[width=0.29\linewidth]{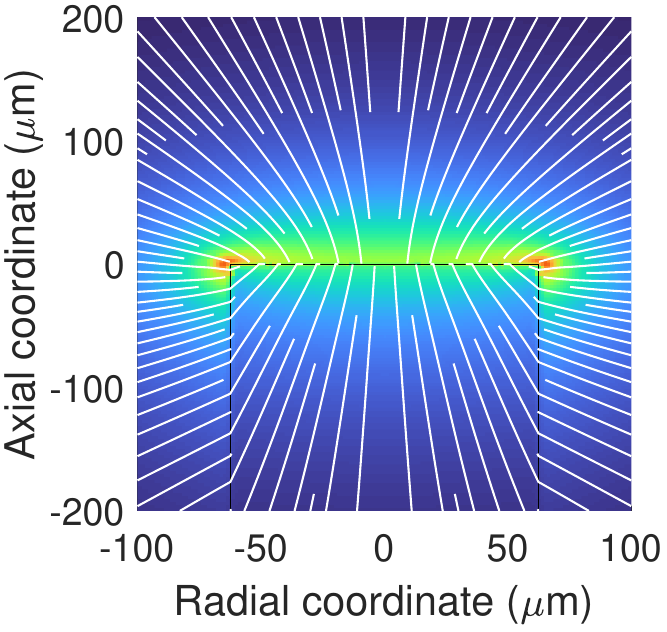} \label{fig:fiber_charge2D_no_shield}}
    \subfloat[]{\includegraphics[width=0.34\linewidth]{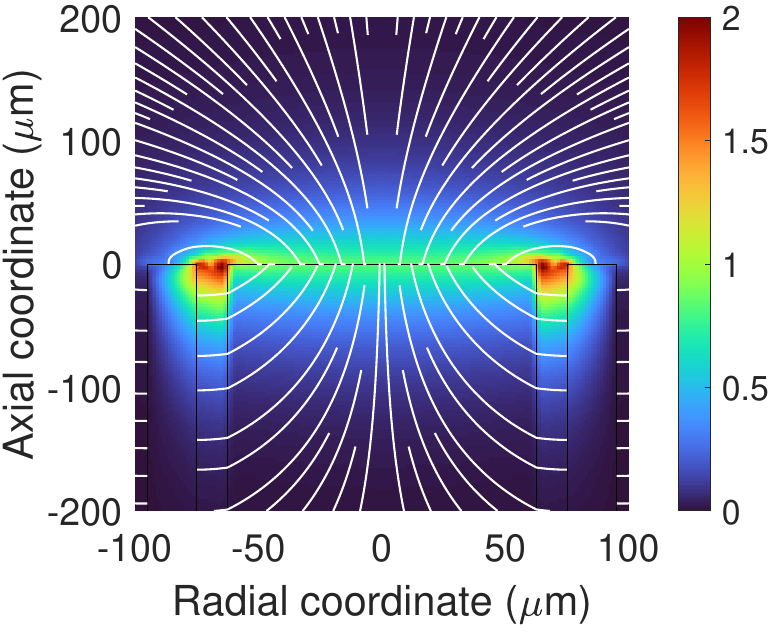} \label{fig:fiber_charge2D_w_shield}}
    \subfloat[]{\includegraphics[width=0.35\linewidth]{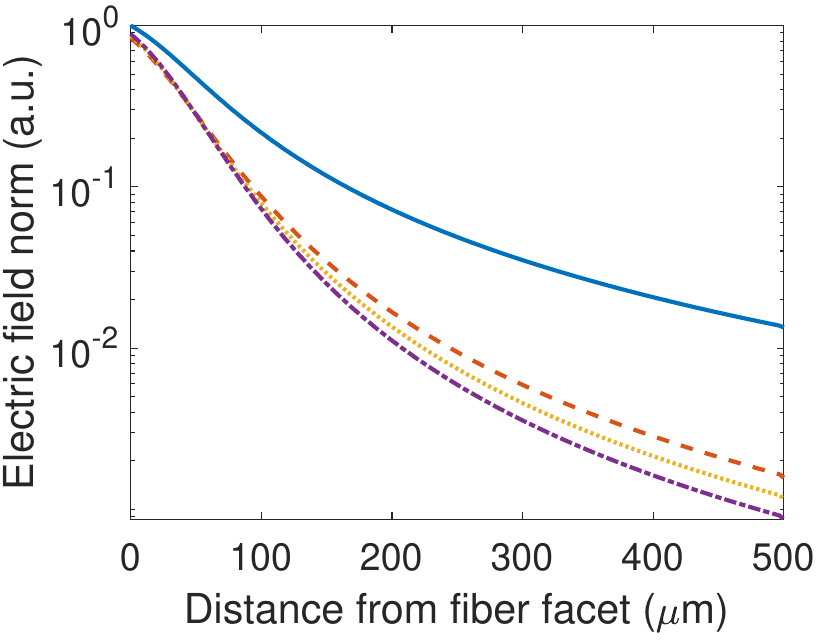} \label{fig:fiber_charge1D}}
    \caption{Heat map distributions of the electric field norm \textbf{(a)} without and \textbf{(b)} with the conductive shield when uniform charge is deposited on the fibre facet. The white lines are guides for the electric field lines. The cross sections of the fibres and the shield are delineated with solid black lines. \textbf{(c)} The electric field norm as a function of the distance from the fibre facet along the axis of the fibre. The plots are without the shield (solid) and with the shield with protrusion of 0 (dashed), 10 (dotted) and 20~$\mum$ (dash-dotted) with respect to the fibre facet.}
    \label{fig:fiber_charge}
\end{figure*}

The primary challenge in the integration of FFPCs is charge deposits on the dielectric fibre surfaces which will disrupt the trapped ions \cite{Ong2020}. Even though the conductive shield introduced in Figure~\ref{fig:Geom} only covers up the side surface of the fibre and the front facet is entirely exposed, due to the proximity of the shield the effect of the deposited charges can be suppressed. 
In \refig{fig:fiber_charge2D_no_shield} and \ref{fig:fiber_charge2D_w_shield} the cross sectional distributions of the electric field norm ($=|{\vec{E}}|$) when uniform charge is deposited on the fibre facet are shown. These numerical simulations are carried out with and without the conductive shield respectively. 
Note that the electric field norm is normalized by the value at the center of the fibre facet without the conductive shield so that the results do not depend on the value of the charge density. 
As can be seen in \refig{fig:fiber_charge2D_w_shield}, the electric field from the deposited charge is diverted towards the shield, which reduces the magnitude of the field at other locations. \refig{fig:fiber_charge1D} shows the electric field norm along the central axis of the fibre as a function of the distance from the fibre facet. It clearly demonstrates the reduction of the field amplitude due to the conductive shield. Modest improvement can be obtained by protruding the shield with respect to the plane of the fibre facet.

\begin{figure}[t]
    \centering
    \subfloat[]{\includegraphics[width = 0.8\linewidth]{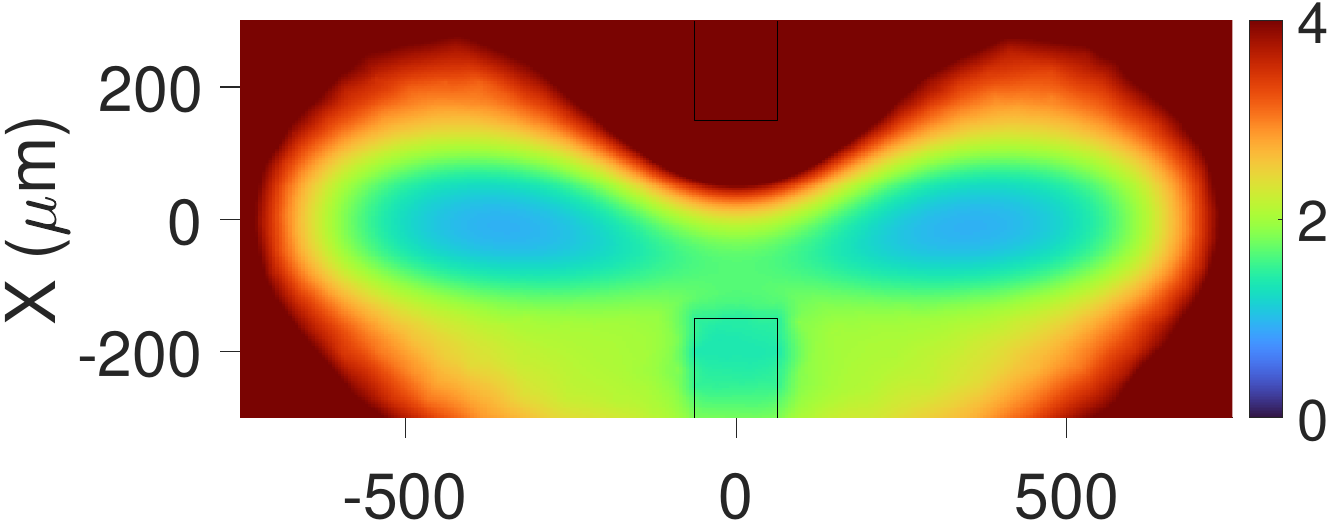}\label{fig:fiber-charge-pseudo-no-shield}}\\
    \subfloat[]{\includegraphics[width = 0.8\linewidth]{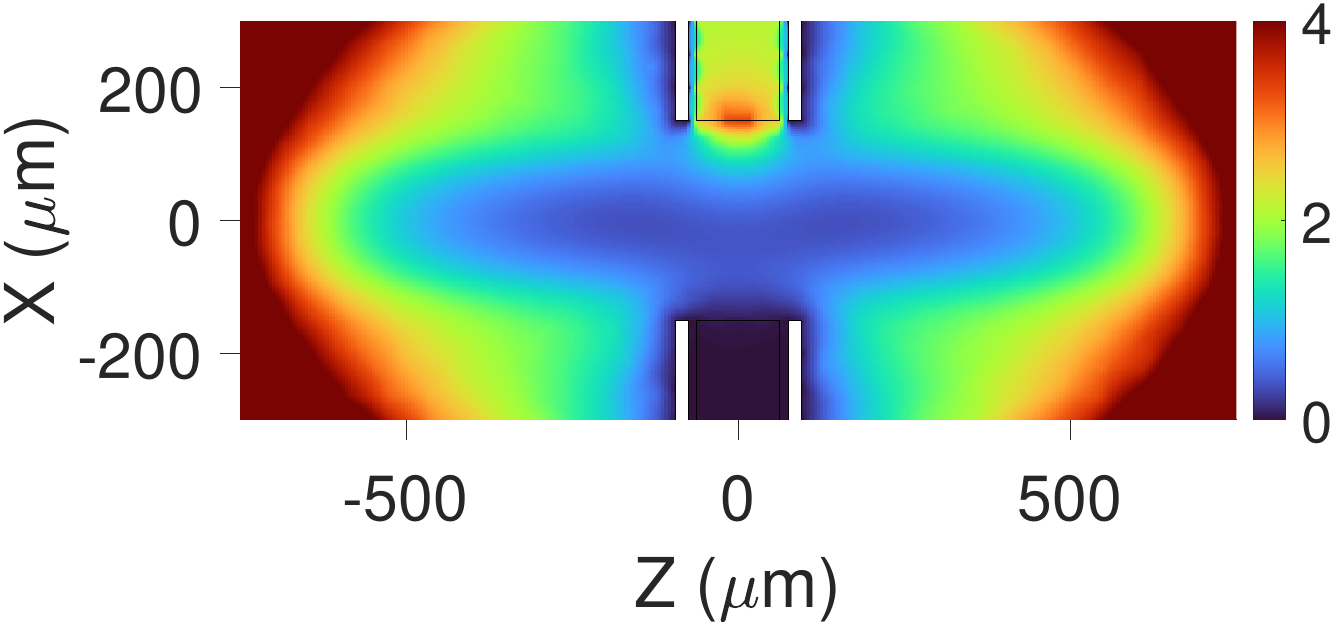}\label{fig:fiber-charge-pseudo-w-shield}}\\
    \subfloat[]{\includegraphics[width = 0.8\linewidth]{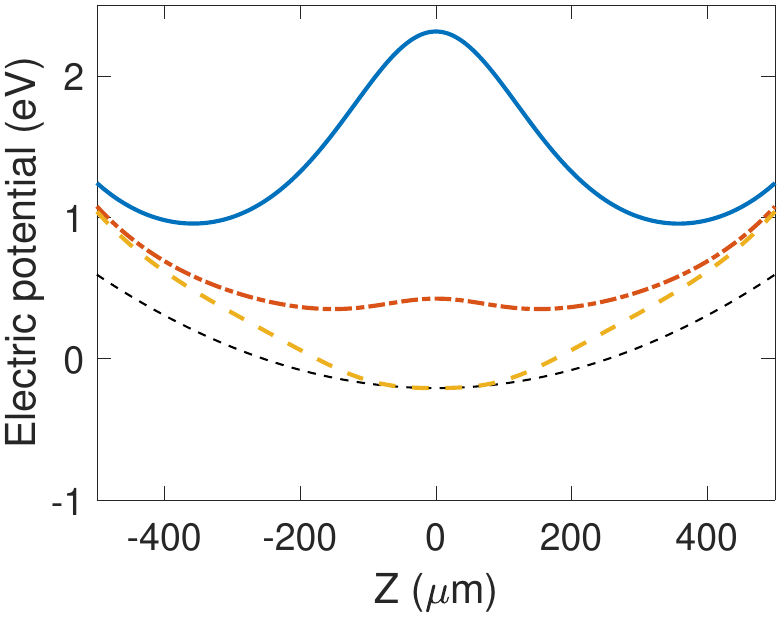}\label{fig:fiber-charge-pseudo-1D}}
    \caption{\textbf{(a-b)}, Cross sectional electric potentials on the XZ plane which include both the pseudopotential and the static potential from the endcap electrodes as well as from stray charges. A charge density of $+10 e/\mum^2$ is applied to the upper fibre. The heat maps are in units of eV. In (a) no shield is included around the fibres. (b) Conductive shields surround both fibres. \textbf{(c)}, The 1D axial potential at the transverse center ($x=y=0$). Blue (solid): without shielding for the fibres. Orange (dash-dotted): With the shield and without any dc compensation voltage. Yellow (dashed): With the shield and dc compensation. The black dashed line is a fit of a harmonic function to the compensated potential around $z=0$.}    
    \label{fig:charge_applied}
\end{figure}

Even though the overall reduction of the electric field shown in \refig{fig:fiber_charge1D} is moderate (-a factor of 3-5 at an ion-fibre distance of 150~$\mum$ depending on the protrusion) it can be shown that the shield is highly useful for nulling the stray electric field from the charge deposits in the actual ion trap. 
We consider simulation of the electric potential in the ion trap shown in \refig{fig:Geom}. Here we set the endcap voltage at 150~V to have an axial secular frequency of 624~kHz. The protrusion of the shield is set to be zero. \refig{fig:charge_applied}~(a-b) show cross sectional views of the total trapping potential $\Phi$ (see Eq.~\refeq{total_pot}) when a surface charge density of $+10~e/\mum^2$ is applied to one of the fibres. This charge density is on the same magnitude as the one reported in \cite{Ong2020}. In the absence of any shielding for the fibres, as a result of excess charge the axial trapping potential is deformed into a double-well with a potential barrier of more than 1~eV in the middle (see \refig{fig:charge_applied}(c)). In this case a single ion would be pushed out of the cavity mode by $\sim360~\mum$ along the trap axis. There is also a relatively small shift for the potential minimum in the $x$ direction. Upon introducing the shields, the potential barrier is significantly reduced to $\sim 0.07$~eV. On the other hand the double-well is still present even though the distance of the well from the center is reduced to 150~$\mum$. Therefore this is still not sufficient for coupling the ion to the cavity. It is, however, found that by applying an additional voltage of -2.4~V on the shield around the fibre on which the charge is deposited and 0.8~V on the other shield, it is possible to almost reverse the distortion of the trapping potential and make it into a single well again. In this case the ion can be placed at the center of the cavity. 
The local secular frequency at $z=0$ is also restored to 626~kHz. As can be seen in \refig{fig:charge_applied}(c), the potential remains to be anharmonic outside the central region. Obviously applying a pair of voltages onto the shields is not enough to completely cancel the effect of the charge deposit across the entire trapping region. More elaborate trap designs which include segmented dc electrodes could be used to address this anharmonicity with more freedom.

\subsection{Motional heating}
\label{subsec:heating}

As has been found in previous studies \cite{Turchette2020, Brownnutt2015}, electric field noise is a major contributor to the heating rate of trapped ions. In this section, we follow the model outlined in ref. \cite{Teller2021} to compute the ion heating rate in the trap geometry shown in \refig{fig:Geom}. We specifically focus on the effects of introducing the FFPC to the ion trap.

The ion gets heated due to noisy electric fields from the environment which couple to its charge \cite{Brownnutt2015}. 
Following the fluctuation dissipation theorem (FDT) \cite{Callen1951}, in a system composed of an ion with surrounding environment, the energy of the ion dissipated in the environment must be balanced by the energy injected into the ion by the environment. When the ion is trapped inside an FFPC, the dielectric mirror coatings on the fibre tips act as dominant lossy environment for the ion. The energy dissipation in the dielectrics in turn creates corresponding electric field noise at the ion due to the FDT \cite{Teller2021}.   

In order to calculate the electric filed noise at the ion using the FDT, it is necessary to simulate the ac electric field in the dielectric layers of the coating caused by the ion's motion. We can numerically simulate the electric field distribution in the dielectric layers at any point of the ion's motion. We refer to $\vec{E}^\zeta_i(\vec{r}) = \vec{E}^a_i(\vec{r}) - \vec{E}^0(\vec{r})$ as the ac electric field amplitude at a given position $\vec{r}$ where $\vec{E^0}(\vec{r})$ is the electric field distribution caused by the ion at its equilibrium position and $\vec{E}^a_i(\vec{r})$ is the electric field distribution when the ion is displaced by an amount $a$ in the direction $i \in \{x,y,z\}$. As per the method used in \cite{Teller2021}, we deduce the electric field noise power spectrum due to the dissipation in the mirror coatings at temperature $T$ using
\begin{equation}
  S_i(\omega) = \frac{4k_BT}{a^2e^2\omega} \epsilon_0 \int_V \epsilon_r(\vec{r}) \tan\delta(\vec{r})\abs{\vec{E}^\zeta_i(\vec{r})}^2 d\vec{r}, \quad i = x, y, z \label{eq:noise_spectrum}
\end{equation}
where $k_B$ is the Boltzman constant, $\epsilon_0$ the permittivity of vacuum, $\omega$ the angular frequency of the ion's secular motion, $V$ the volume of the dielectric material, $\epsilon_r$ and $\tan\delta$ the relative permittivity and loss tangent of the material respectively. Note that when the ion's oscillation amplitude is small ($a \ll 1$), $\abs{\vec{E}^\zeta_i} \propto a$ is satisfied. Therefore in this regime, the electric field noise spectrum \refeq{eq:noise_spectrum} becomes independent of $a$.

The high reflectivity mirror coating on the fibre facet is modelled as a stack of SiO$_2$ and Ta$_2$O$_5$ coating pairs, layered 20 times. We use relative permittivities of 3.75 and 22, and loss tangents of $1.3\cdot10^{-3}$ and $7.0\cdot10^{-3}$ for SiO$_2$ and Ta$_2$O$_5$, respectively \cite{Teller2021}. For computational simplicity of the finite element modelling method that is employed, each single coating layer is modelled as 250 nm thick (instead of the nominal $\lambda/4$ thickness, where $\lambda$ is the operational wavelength of the cavity). We infer the motional heating rate in each mode of the ion's motion in terms of number of quanta using \cite{Brownnutt2015}
\begin{equation}
    \dot{n}_i = \frac{e^2}{4m\hbar\omega}S_i(\omega_i), \quad i = x, y, z, \label{eq:heating_rate}
\end{equation}
where $\omega_i$ specifies the motional secular frequency.
The secular frequencies used in the simulation are $2\pi\cdot$1 MHz for the axial mode and $2\pi\cdot$3 MHz for the radial modes, and the reader can estimate the heating rate for any secular frequency by making the appropriate quadratic scaling.

The effect of the shielding is analogous to the one described in Section~\ref{subsec:charging}. The shield distracts and absorbs a part of the electric field generated by the ion. As a result, the proportion of the electric field in the dielectric coatings is reduced in comparison with the case without the shielding. 

\begin{figure*}[t!]
       \captionsetup[subfloat]{position=top}
       \subfloat[]{\includegraphics[width=0.29\linewidth]{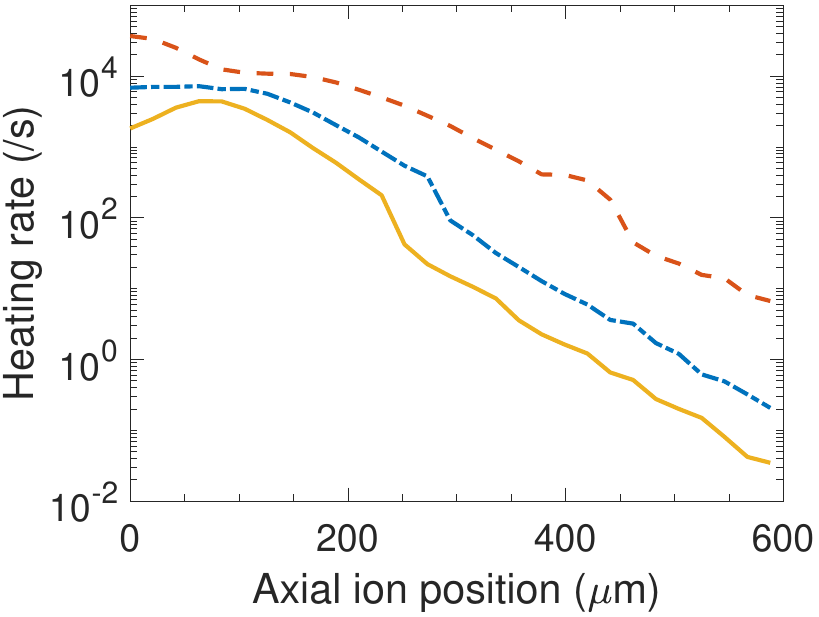}
       \label{fig:heating_rate_ax}} 
       \subfloat[]{\includegraphics[width=0.29\linewidth]{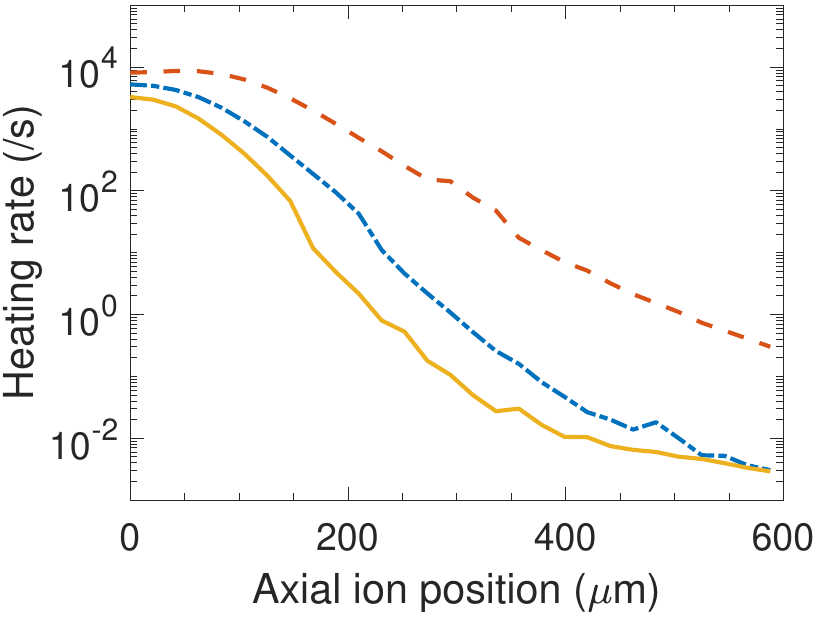}
       \label{fig:heating_rate_radx}}
        \subfloat[]{\includegraphics[width=0.29\linewidth]{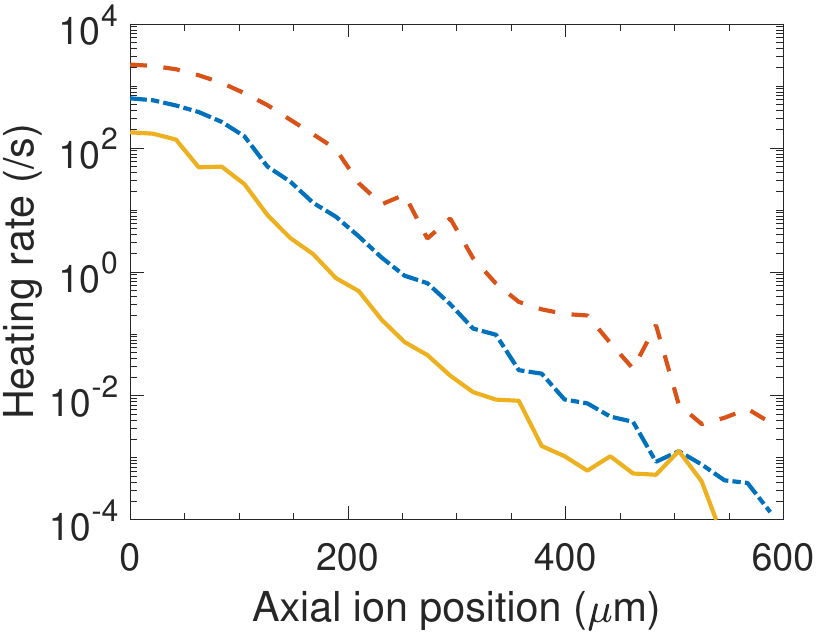}
        \label{fig:heating_rate_rady}}
        \hspace{3pt}
        \subfloat[]{\includegraphics[width=0.08\linewidth]{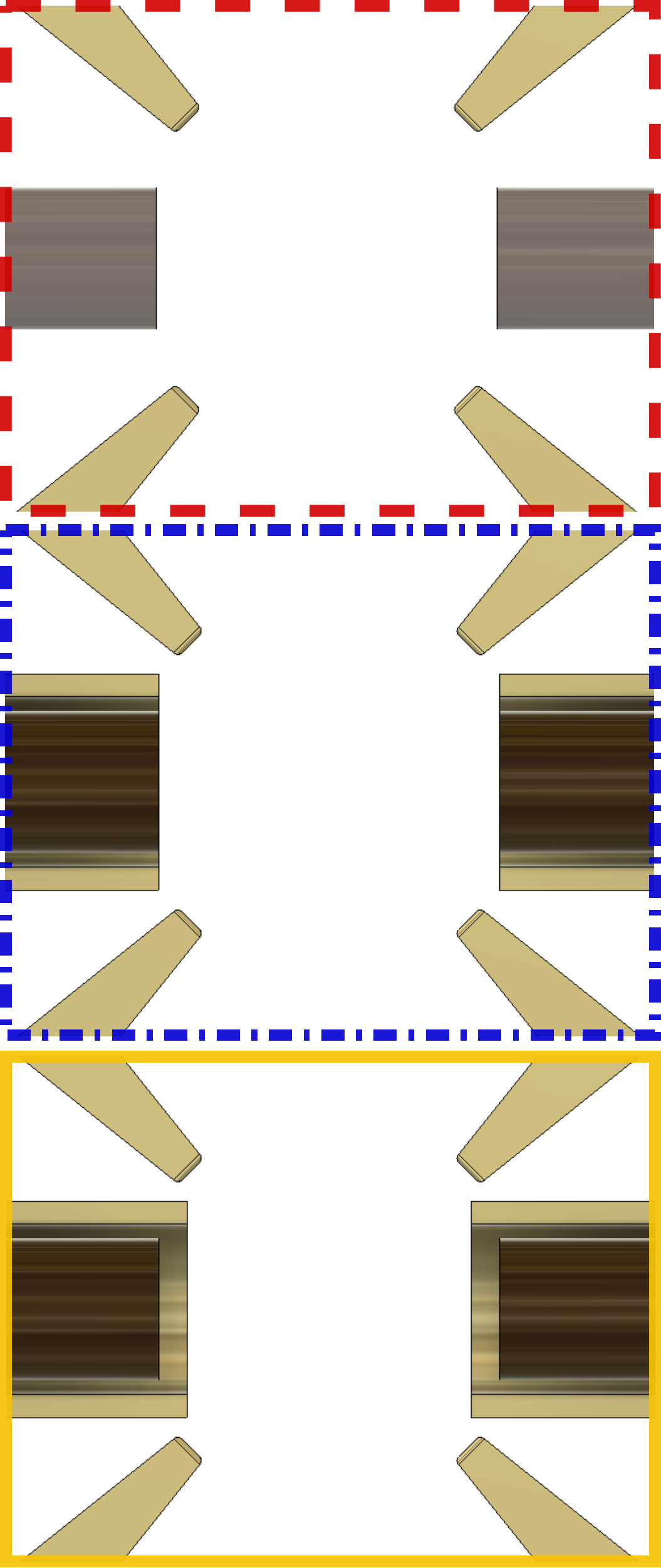}
       \label{fig:heating_rate_cartoons}} 
    \caption{Simulated motional heating rates for \textbf{(a)} the axial (along the z-axis), \textbf{(b)} radial (x-axis), and \textbf{(c)} radial (y-axis) motional modes, plotted as functions of the ion's axial equilibrium position. The dashed lines correspond to the cases where no shielding is included. The dash-dotted lines are with the shielding with no protrusion of the shield. The solid lines correspond to the cases where the shield is protruded by 25~$\mum$ whilst keeping the cavity length the same. The drawings in \textbf{(d)} depicts these three configurations. Note that the unshielded case has large numerical errors beyond $\sim250~\mum$.}
    \label{fig:heating-rate}
\end{figure*}

The simulation results are presented in \refig{fig:heating-rate}. We have repeated the simulation while the equilibrium position of the ion is varied along the trap axis and calculated the heating rates for the three orthogonal motional modes at each equilibrium position based on \refeq{eq:noise_spectrum} and \refeq{eq:heating_rate}. The origin of the axial ion position in the figures corresponds to the center of the cavity mode. The results show a significant heating rate contribution from the dielectric coatings when the ion is in the cavity mode.  
When the ion is near the cavity mode, the total heating rate is about 5 times smaller with the shield than without the shield. As the ion is translated away from the cavity mode along the trap axis, we find
the heating rate drops faster with the shield than without the shield. We also note that the heating
rates drop to below 1 quanta/s for the radial modes when the ion is translated by about 300~$\mum$.
We find that we can further reduce the heating rate by protruding the shields similar to the experimental setup in \cite{Takahashi:20}. Here we keep the cavity length constant and protrude each shield by 25~$\mum$ towards the trap centre as can be seen in \refig{fig:heating_rate_cartoons}. 

Fast two-qubit gates have been demonstrated in systems with heating rates as high as 100 quanta/s \cite{Schafer2018}. While heating rates as low as possible are desired, the upper bound should be set according to the desired function of the qubits, whether they serve as memory or as gate qubits. 
While the heating rates of $10^2-10^3$ presented here when the ion is in the cavity are large for constant operation, the ion could be shuttled away from the cavity mode for significantly reduced heating rates. This scheme is promising because information is stored in the internal states of the ion while the ion is in the cavity mode. Thus the information remains intact during and after the shuttling. The information-carrying ion can be sympathetically cooled before the information is transferred to other ion via motional coupling. The shuttling can also be operated at a fast rate as has been demonstrated in \cite{Walther2012} to reduce the overhead in the entanglement generation rate.

\begin{figure}
    \centering
    \includegraphics[width=0.8\linewidth]{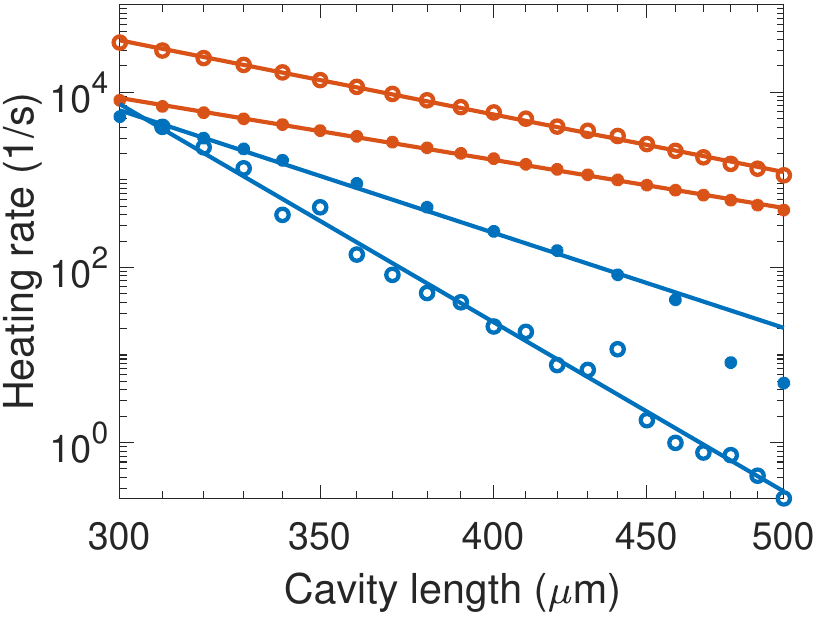}
    \caption{Simulated heating rates of the cavity axis mode (dots) and the trap axis mode (circles) versus the cavity length. The red data correspond to the case where the FFPC shield is not included and the blue data corresponds to the case where the shield is included. The lines are fits for extracting the exponent $\alpha$. See main text for more details.  }
    \label{fig:heating-rate-v-cav-length}
\end{figure}

Encouraged by the above results, we further analyze the dependence of the heating rate on the cavity length. The results are shown in \refig{fig:heating-rate-v-cav-length}. The red circles and lines correspond to the simulation results and fits when no shield is employed. The simulation is carried out as the fibres are retracted in the $x$ direction away from the trap center symmetrically. In the figure, the heating rates of the cavity axis ($x$) and trap axis ($z$) modes are presented. They are fitted well by a function $\propto d^{-\alpha}$ where $d$ is the distance between the ion and the fibre facet with the cavity length given by $2d$. The exponents are obtained as $\alpha = 5.7$ ($x$) and $6.8$ ($z$) respectively. Note that even without the shields around the fibres, the rf electrodes in front of the fibres act like partial shields as the fibres are retracted. We have confirmed this by calculating the heating rates without the rf electrodes (not shown) and obtained exponents of 4.4 ($x$) and 5.7 ($z$) respectively. The difference in the exponents in these results with and without the rf electrodes show that the presence of conductive materials in the proximity of the dielectrics can change the distance scaling of the heating rate. This effect is more prominent when the shields around the fibres are introduced (see blue data in \refig{fig:heating-rate-v-cav-length}). In this case, the fibres are retracted within the shields while the positions of the shields are kept fixed. The exponents are extracted from the fits to be $11.2$ ($x$) and $19.9$ ($z$). Therefore conductive shields around the FFPC alter the distance scaling of the heating rates drastically, enhancing the suppression of the heating rates as the cavity length is increased. Combined with the shuttling scheme discussed earlier, this finding opens up a new possibility: With the conductive shields, a moderate increase of the cavity length provides further reduction of the heating rates by orders of magnitude, which may bring the heating rate to a level acceptable for the operation of quantum processors (see also the discussion in Section~\ref{sec:discussion}).             

\section{The price of the shield: deformation of the trapping potential}
\label{sec:price}

We have seen in the previous sections that shielding the FFPC with a conductor is highly beneficial for mitigating both static charge and motional heating around the FFPC. Here we examine the effects of this conductive shield on the trapping potential. We find that presence of the shield heavily alters the trapping potential when the rf electrodes are driven in the conventional, and easier to implement, manner with a single-rf signal, rendering the use of the shield counter-productive in this case. We find that an elegant remedy to this is to use the dual-rf drive. 
In the following we compare simulations of the pseudopotential in three different cases. Their simulated conditions are listed in Table~\ref{tab:cases} and they are referred to as the case I, II and III. The differences between these cases are whether the conductive shield is present (I vs II and III), and whether the single or dual-rf drive is used (I and II vs III). In all the cases, the FFPC is present, and the shield is electrically grounded when included in the simulation.

\begin{table}[htb]
    \centering
    \begin{tabular}{|c|c|c|c|} 
     \hline
     Case & FFPC included? & Shield included?& Rf drive\\
     \hline
     I & Yes & No & Single \\ 
     \hline
     II & Yes & Yes & Single \\ 
     \hline
     III & Yes & Yes & Dual \\ 
     \hline
    \end{tabular}
    \caption{Simulated conditions for three different cases.}
    \label{tab:cases}
\end{table}

\begin{figure*}[!ht]
    \centering
        \subfloat[]{\includegraphics[width=0.27\linewidth]{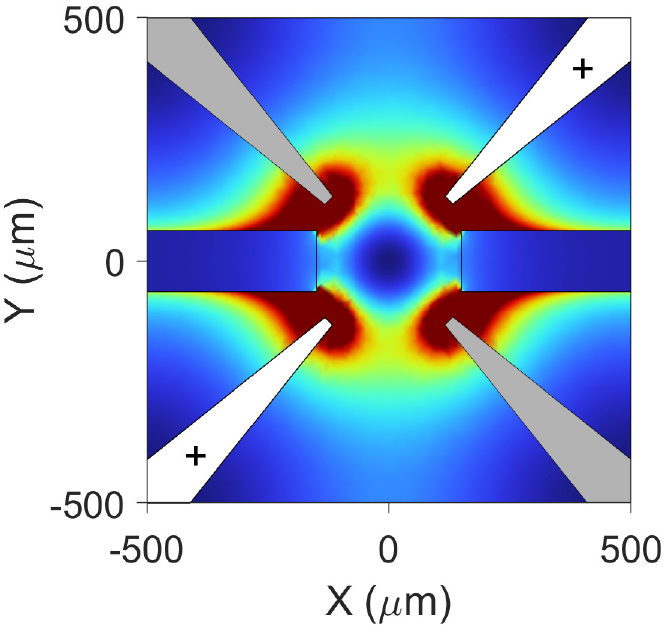}
        \label{fig:radial-potential-no-shield}}
        \hspace{10pt}
        \subfloat[]{\includegraphics[width=0.27\linewidth]{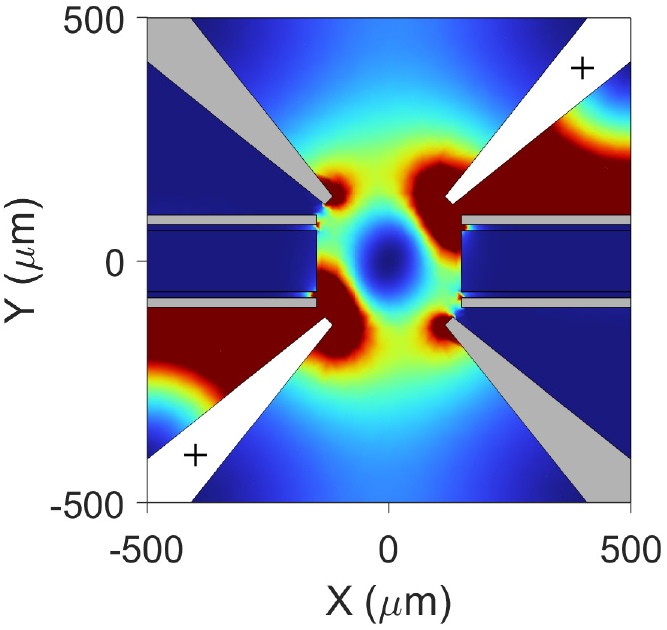}
        \label{fig:radial-potential-w-shield-single}}
        \hspace{10pt}
        \subfloat[]{\includegraphics[width=0.32\linewidth]{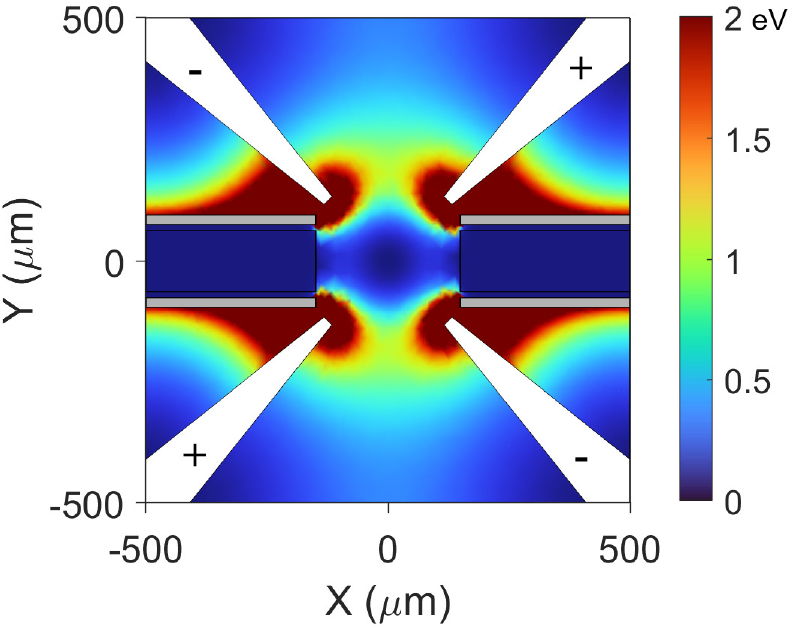}
        \label{fig:radial-potential-w-shield-dual}}
    \\
        \subfloat[]{\includegraphics[width=0.27\linewidth]{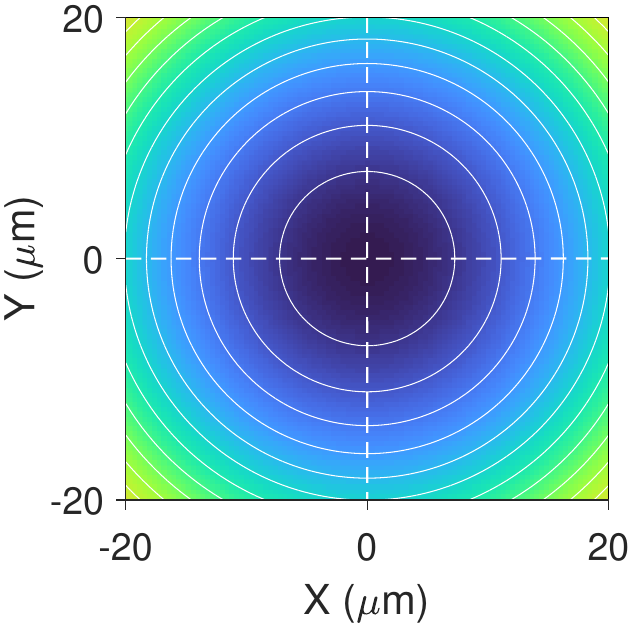}
        \label{fig:radial-potential-no-shield-cont}}
        \hspace{10pt}
        \subfloat[]{\includegraphics[width=0.27\linewidth]{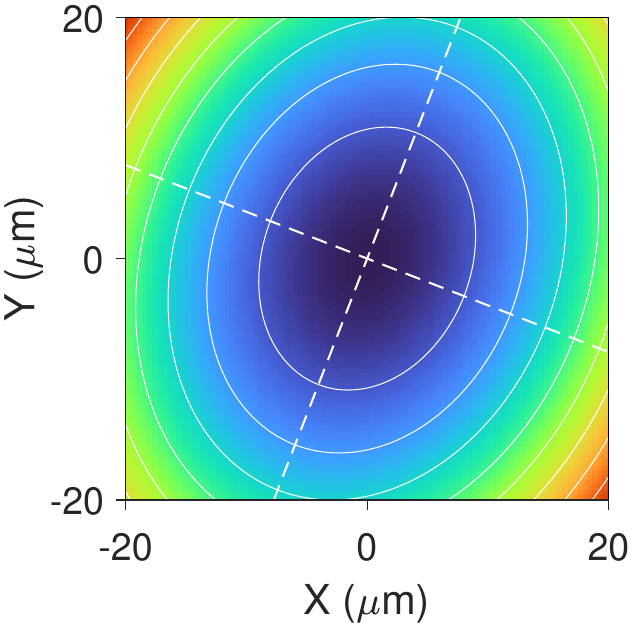}
        \label{fig:radial-potential-w-shield-single-cont}}
        \hspace{10pt}
        \subfloat[]{\includegraphics[width=0.334\linewidth]{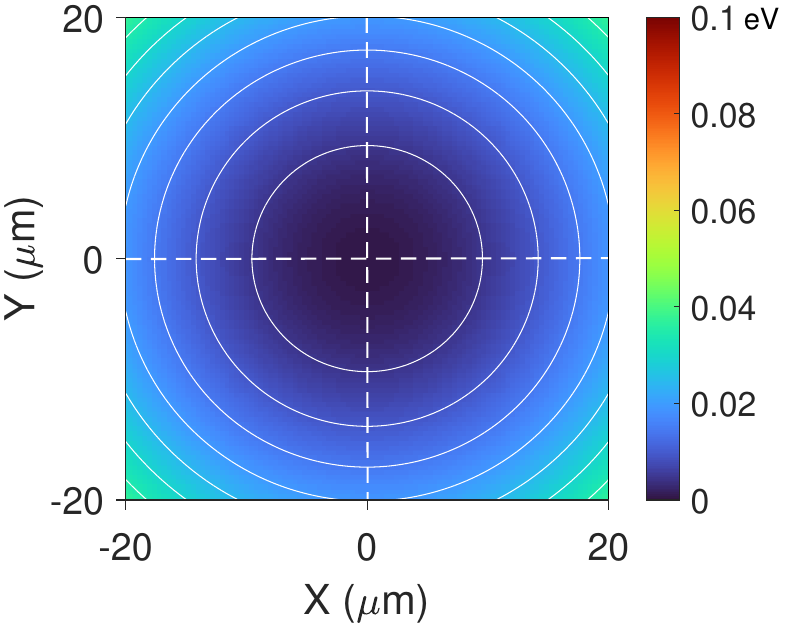}
        \label{fig:radial-potential-w-shield-dual-cont}}
    \caption{\textbf{(a-c)}, Cross-sectional plots of the radial pseudopotentials at around the center of the FFPC (a) without the shield and single-rf drive, (b) with the shield and single-rf drive, and (c) with the shield and dual-rf drive. Electrically grounded parts are colored gray, and the polarities of the rf voltage are indicated by +/- symbols. \textbf{(d-f)} Close-up views of the same potentials as in (a-c). The white solid and dashed lines indicate equipotential contours and the principal axes respectively.}
    \label{fig:radial-potential}
\end{figure*}

\subsection{Radial potential}
\label{sec:rad_potential}

First we study the pseudopotential in the radial plane. \refig{fig:radial-potential} shows cross-sectional plots of the pseudopotentials in the $xy$ plane in the case I (\subref*{fig:radial-potential-no-shield} and \subref*{fig:radial-potential-no-shield-cont}), II (\subref*{fig:radial-potential-w-shield-single} and \subref*{fig:radial-potential-w-shield-single-cont}), and III (\subref*{fig:radial-potential-w-shield-dual} and \subref*{fig:radial-potential-w-shield-dual-cont}). 
As can be seen in \refig{fig:radial-potential-no-shield}, when the shield is not employed, the FFPC itself does not incur substantial deformation to the radial pseudopotential. As is shown by the contour in \refig{fig:radial-potential-no-shield-cont}, the symmetry of the potential is maintained at the center and secular frequencies also remain almost the same as the ones without an FFPC.
On the other hand, when the shield is incorporated and the single-rf drive is employed, the radial pseudopotential is deformed as shown in \refig{fig:radial-potential-w-shield-single} and \refig{fig:radial-potential-w-shield-single-cont}. It is evident in the contour plot in \refig{fig:radial-potential-w-shield-single-cont} that the potential is no longer symmetric but is elliptic, of which principal axes are also tilted by $\sim 21^\circ$.
The secular frequencies along these principal axes are $3.7$ and $3.1$~MHz respectively. This deformation is caused because the presence of the grounded shield breaks the symmetry of the electric field distribution amongst the electrodes.
The symmetry can be restored by using the dual-rf drive as shown in \refig{fig:radial-potential-w-shield-dual} and \refig{fig:radial-potential-w-shield-dual-cont}. Nonetheless  
small ellipticity remains as the secular frequencies are also lowered to $2.3$ and $2.6$~MHz along the two principal axes.

Even though there are some deformations due to the shield most notably in the case of the single-rf drive as seen above, these deformations in the radial potentials alone are not critical. A small disparity of the secular frequencies is sometimes rather favoured. In addition, if necessary, the modified secular frequencies and tilt of the principal axes can be tuned by the rf voltages and the Mathieu `a' parameter of the ion trap. The main interference of the shield with the rf potential is exhibited along the axial direction as discussed next.

\subsection{Axial potential}
\label{sec:axial_potential}

\begin{figure*}[!t]
    \centering
     \subfloat[]{\includegraphics[width=0.32\linewidth]{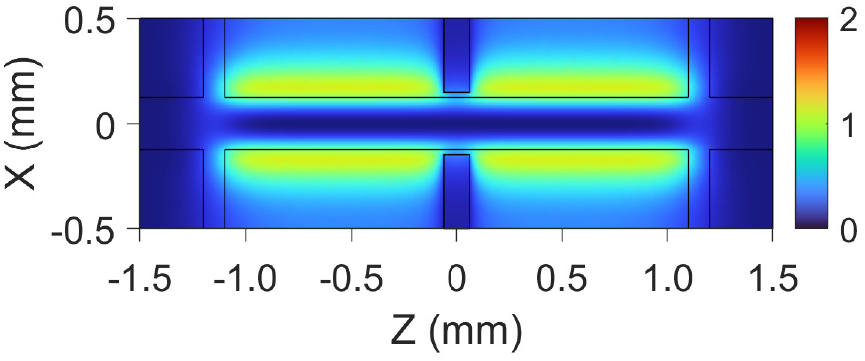}
     \label{axial_pot_2D_no_shield}}
     \subfloat[]{\includegraphics[width=0.32\linewidth]{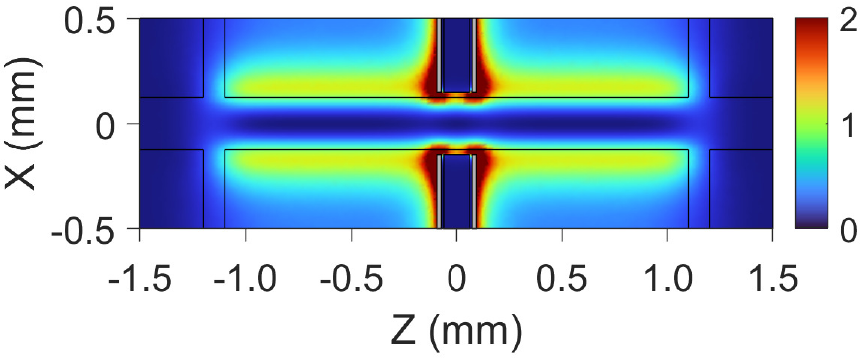}
     \label{axial_pot_2D_w_shield_asym}}
     \subfloat[]{\includegraphics[width=0.32\linewidth]{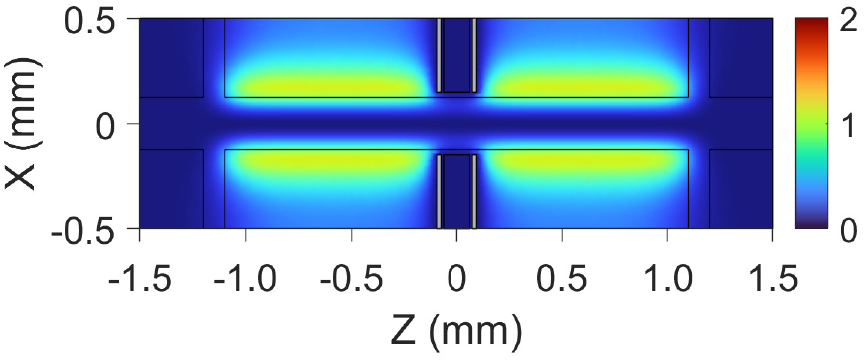}
     \label{axial_pot_2D_w_shield_sym}}
     \\
     \captionsetup[subfloat]{captionskip=0pt}
     \subfloat[]{\includegraphics[width=0.32\linewidth]{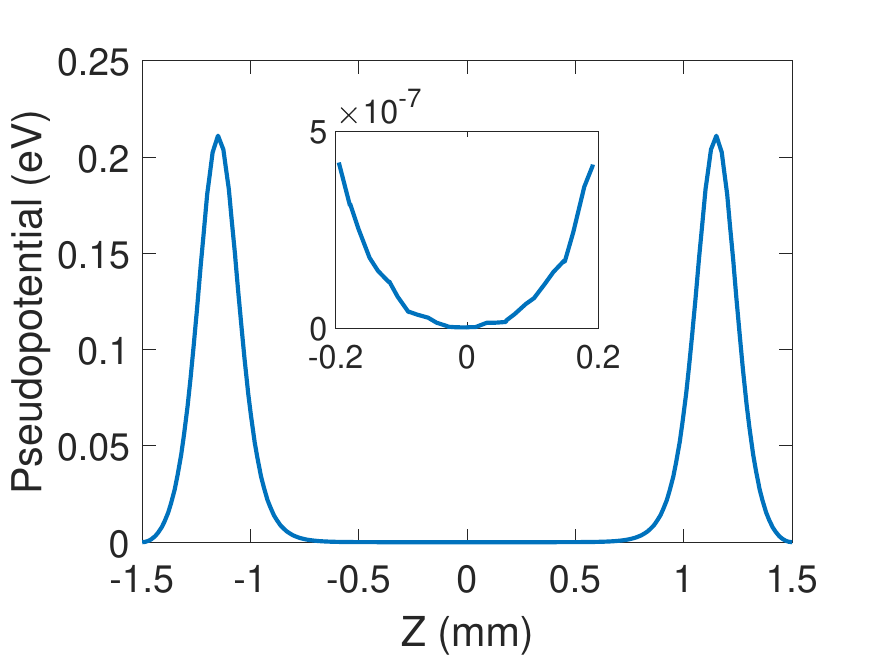}\label{axial_pot_1D_no_shield}}
     \subfloat[]{\includegraphics[width=0.32\linewidth]{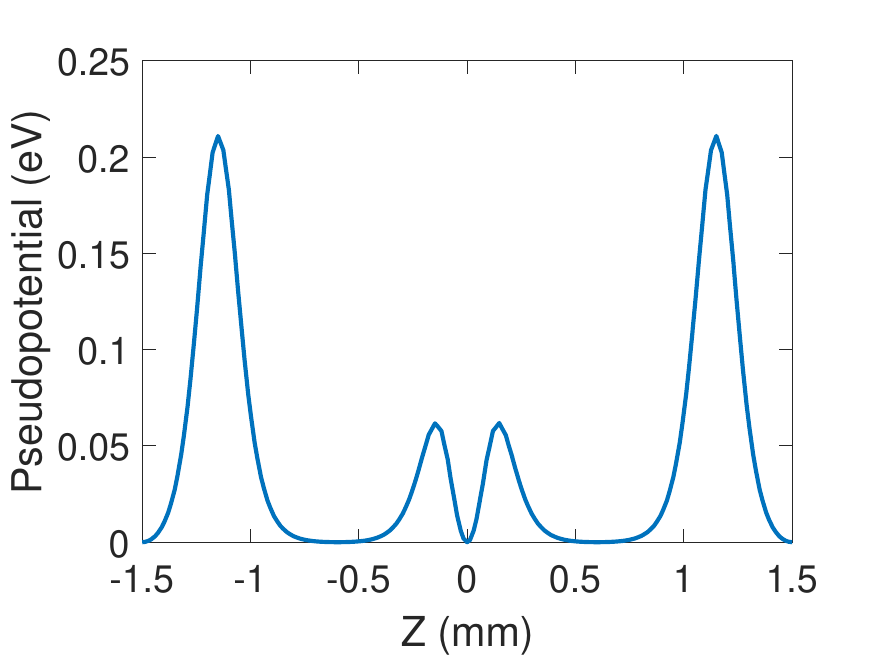}\label{axial_pot_1D_w_shield_asym}}
     \subfloat[]{\includegraphics[width=0.32\linewidth]{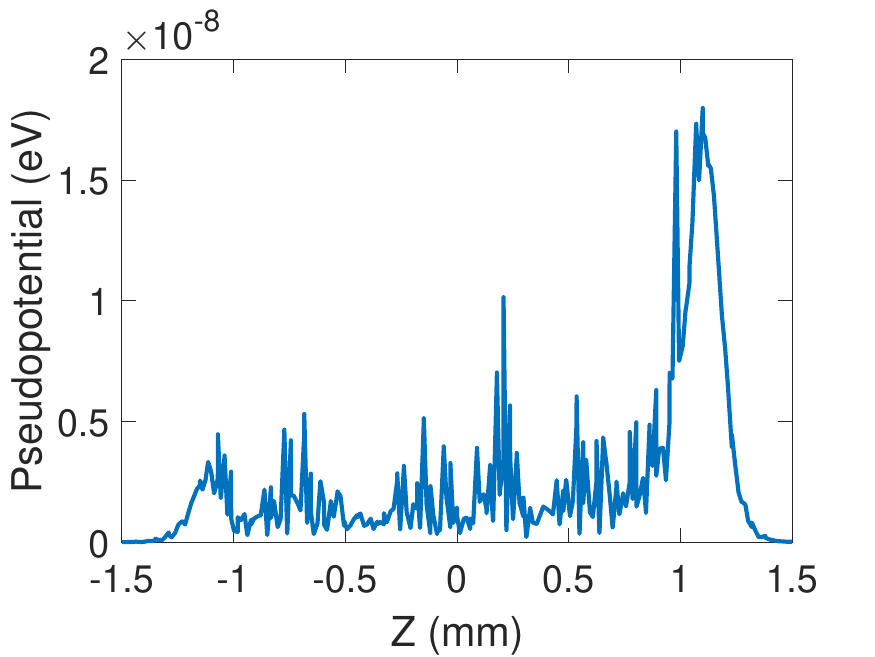}\label{axial_pot_1D_w_shield_sym}}
    \caption{\textit{Top}: 2D plots of the pseudopotential in the $xz$ plane ($y = 0$). \textit{Bottom}: 1D plots of the pseudopotential along the trap axis ($x=y=0$). The simulations are done for the case I (a, d), II (b, e) and III (c, f). The 2D heatmaps are in the units of eV. The inset in (d) shows a close-up around $z=0$.}
    \label{fig:axial-potential}
\end{figure*}

\begin{figure}[thb]
    \centering
    \includegraphics[width = 0.8\linewidth]{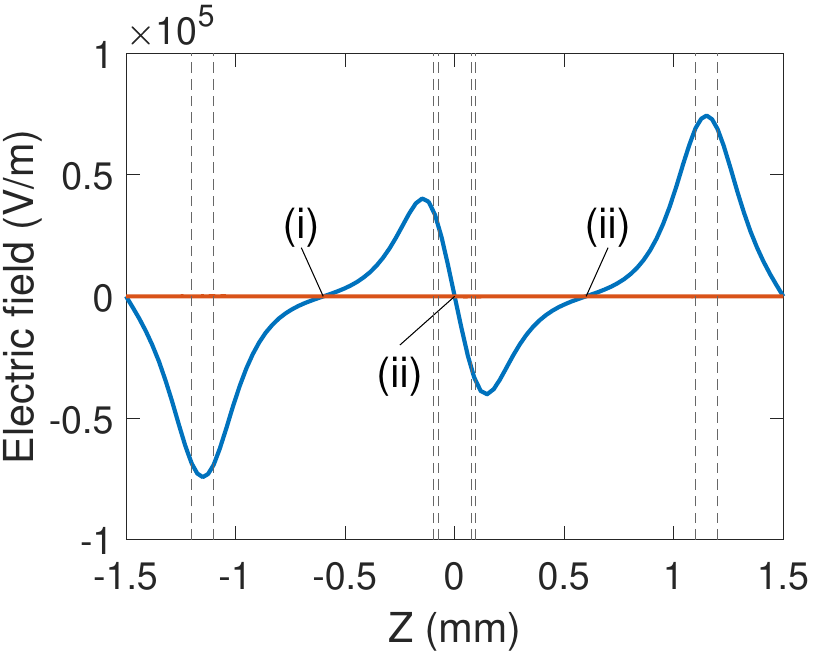}
    \caption{Vectorial components of the electric field amplitude on the trap axis ($=\vec{E}_0(0, 0, z)$ in \refeq{E_rf}) simulated in the case II. The z component (=$E_{0z}$) is shown as a blue solid curve. The x and y components (=$E_{0x/y}$) are seen degenerate in this scale and shown as the orange line. The dashed vertical lines indicate the boundaries of the conductive shield and the rf/endcap electrodes. The three labels (i)-(iii) point to the locations where the electric field becomes exactly zero.}
    \label{fig:Efield-components}
\end{figure}

\refig{fig:axial-potential} shows two-dimensional cuts of the pseudopotentials in the $xz$ plane ($y=0$) as well as one-dimensional plots along the trap axis ($x=y=0$) for the cases I, II and III. 
Once again the presence of the fibre alone does not create a notable deformation to the pseudopotential (\refig{axial_pot_2D_no_shield} and \subref*{axial_pot_1D_no_shield}). The increase of the pseudopotential towards the outer edges of the rf electrodes is a common feature when the single-rf drive is employed.
On the other hand the pseudopotential is significantly altered in the case II with the introduction of the conductive shield (\refig{axial_pot_2D_w_shield_asym} and \subref*{axial_pot_1D_w_shield_asym}). Additional local maxima of the pseudopotential form around the front facets of the shield. As a result, a new potential well with barriers as high as 0.06~eV is formed around $z=0$ along the axial direction (\refig{axial_pot_1D_w_shield_asym}). \refig{fig:Efield-components} shows the vectorial components of the electric field on the trap axis simulated in the case II. As is shown, the non-zero pseudopotential is entirely comprised of the $E_z$ component. The magnitude of $E_z$ increases around the locations where the translational continuity of the linear trap is disrupted by the shield or the gap between the endcap and rf electrodes. The latter is also present in \refig{axial_pot_1D_no_shield}.
These features can be completely suppressed by using the dual-rf drive as seen in \refig{axial_pot_2D_w_shield_sym} and \subref*{axial_pot_1D_w_shield_sym}. In this case, the pseudopotential is virtually zero along the trap axis, only limited by the numerical errors of the simulation.        

In a linear ion trap, the pseudopotential should be ideally null along the trap axis. This is the feature that enables linear ion traps to stably maintain a string of ions along the axis. Strictly speaking, however, this can not be realized with the single-rf drive in general as long as the length of the linear ion trap is finite. When the single-rf drive is employed, disruption of the translation symmetry accompanies miscancellation of the rf electric field at the corresponding positions on the trap axis. Without the conductive shield around the fibres as in the case I, this disruption only happens at the boundaries between the rf and endcap electrodes which are far from the central trapping region. As a result the residual rf electric field at the center of the trap remains at a negligible level (see the inset of \refig{axial_pot_1D_no_shield}). By contrast, in the case II with the conductive shield around the fibres, the disruption happens close to the center where ions are expected to be trapped. The created additional potential barriers essentially isolates the central region of the trap from the rest. As can be seen in \refig{fig:Efield-components}, within the trap there are only three separated locations (denoted (i)-(iii) in the figure) on the trap axis where the electric field becomes exactly zero. Therefore the entire trap is segmented into three trapping regions in each of which there is only a single rf null. Away from these rf nulls, trapped ions would suffer from excessive micromotion. These modifications in the axial potential in the case II are obviously unfavored and defy the very purpose of coupling an ion in a linear string to the cavity. Replacing the single-rf drive with the dual-rf drive restores the cancellation of the rf field on the trap axis in the case III. In Section \ref{sec:trap-symmetry} we will analytically study the mechanism behind the cancellation of the rf field and derive the general conditions for the trap symmetry and rf-drive to be robust against disruptive elements such as shielded fibres.

\subsection{Effects of misalignments}
\label{sec:misalignment}

In the previous section, we found that the dual-rf drive is robust against disruptions that break the translational symmetry in the linear trap. However the placement of the fibres and shields itself was still symmetric in the transverse plane. In this section we examine how the potential is modified when there is relative misalignment between the opposing fibres, hence the symmetry of the FFPC in the transverse plane is violated.

\begin{figure}[thb]
    \centering
    \subfloat[]{
        \def\svgwidth{0.2\linewidth}
\begingroup%
  \makeatletter%
  \providecommand\color[2][]{%
    \errmessage{(Inkscape) Color is used for the text in Inkscape, but the package 'color.sty' is not loaded}%
    \renewcommand\color[2][]{}%
  }%
  \providecommand\transparent[1]{%
    \errmessage{(Inkscape) Transparency is used (non-zero) for the text in Inkscape, but the package 'transparent.sty' is not loaded}%
    \renewcommand\transparent[1]{}%
  }%
  \providecommand\rotatebox[2]{#2}%
  \newcommand*\fsize{\dimexpr\f@size pt\relax}%
  \newcommand*\lineheight[1]{\fontsize{\fsize}{#1\fsize}\selectfont}%
  \ifx\svgwidth\undefined%
    \setlength{\unitlength}{84.37635707bp}%
    \ifx\svgscale\undefined%
      \relax%
    \else%
      \setlength{\unitlength}{\unitlength * \real{\svgscale}}%
    \fi%
  \else%
    \setlength{\unitlength}{\svgwidth}%
  \fi%
  \global\let\svgwidth\undefined%
  \global\let\svgscale\undefined%
  \makeatother%
  \begin{picture}(1,2.15584495)%
    \lineheight{1}%
    \setlength\tabcolsep{0pt}%
    \put(0,0){\includegraphics[width=\unitlength,page=1]{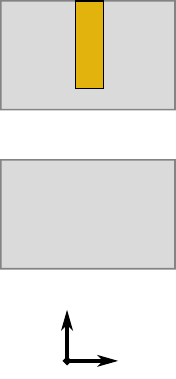}}%
    \put(0.27621638,0.37089504){\makebox(0,0)[lt]{\lineheight{1.25}\smash{\begin{tabular}[t]{l}x\end{tabular}}}}%
    \put(0.63455523,0.00000001){\makebox(0,0)[lt]{\lineheight{1.25}\smash{\begin{tabular}[t]{l}z\end{tabular}}}}%
    \put(0,0){\includegraphics[width=\unitlength,page=2]{Figures/misalignments/misalignement-cartoons-z.pdf}}%
    \put(0.68394235,0.8887391){\makebox(0,0)[lt]{\lineheight{1.25}\smash{\begin{tabular}[t]{l}5$\mum$\end{tabular}}}}%
  \end{picture}%
\endgroup%

        \hspace{10pt}
        \includegraphics[width = 0.74\linewidth]{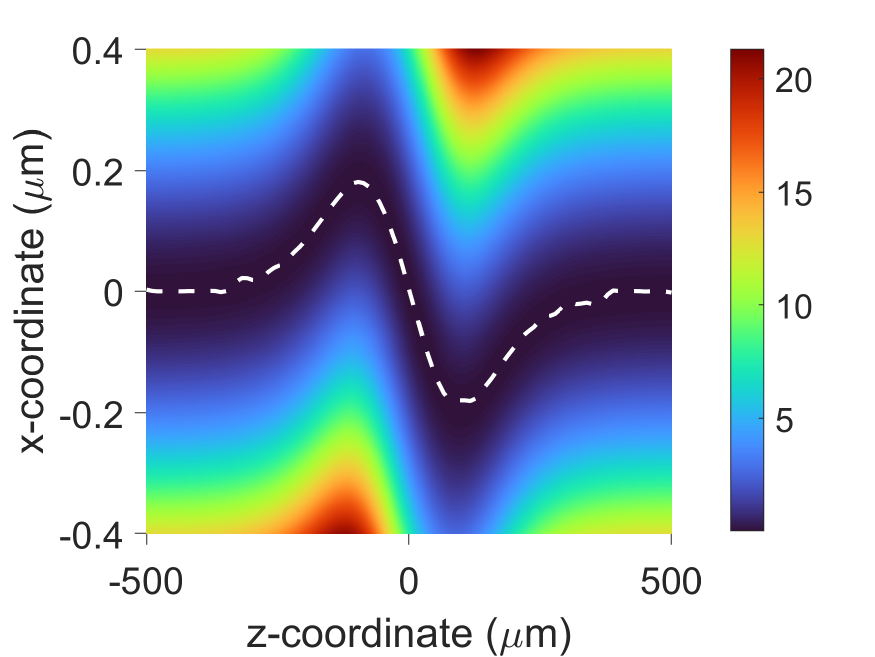}
        \label{fig:ax-offset-potmin}
    }
    \\
    \centering
    \subfloat[]{
        \def\svgwidth{0.2\linewidth}
\begingroup%
  \makeatletter%
  \providecommand\color[2][]{%
    \errmessage{(Inkscape) Color is used for the text in Inkscape, but the package 'color.sty' is not loaded}%
    \renewcommand\color[2][]{}%
  }%
  \providecommand\transparent[1]{%
    \errmessage{(Inkscape) Transparency is used (non-zero) for the text in Inkscape, but the package 'transparent.sty' is not loaded}%
    \renewcommand\transparent[1]{}%
  }%
  \providecommand\rotatebox[2]{#2}%
  \newcommand*\fsize{\dimexpr\f@size pt\relax}%
  \newcommand*\lineheight[1]{\fontsize{\fsize}{#1\fsize}\selectfont}%
  \ifx\svgwidth\undefined%
    \setlength{\unitlength}{84.37635707bp}%
    \ifx\svgscale\undefined%
      \relax%
    \else%
      \setlength{\unitlength}{\unitlength * \real{\svgscale}}%
    \fi%
  \else%
    \setlength{\unitlength}{\svgwidth}%
  \fi%
  \global\let\svgwidth\undefined%
  \global\let\svgscale\undefined%
  \makeatother%
  \begin{picture}(1,2.15584495)%
    \lineheight{1}%
    \setlength\tabcolsep{0pt}%
    \put(0,0){\includegraphics[width=\unitlength,page=1]{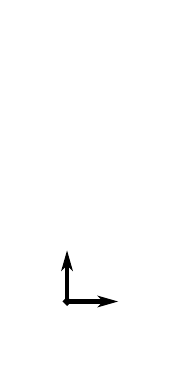}}%
    \put(0.68890858,0.37279494){\makebox(0,0)[lt]{\lineheight{1.25}\smash{\begin{tabular}[t]{l}x\end{tabular}}}}%
    \put(0.26884787,0.73776149){\makebox(0,0)[lt]{\lineheight{1.25}\smash{\begin{tabular}[t]{l}y\end{tabular}}}}%
    \put(0,0){\includegraphics[width=\unitlength,page=2]{Figures/misalignments/misalignement-cartoons-y-rotated_v2.pdf}}%
    \put(-0.062041,1.20095658){\makebox(0,0)[lt]{\lineheight{1.25}\smash{\begin{tabular}[t]{l}5 $\mu$m\end{tabular}}}}%
    \put(0,0){\includegraphics[width=\unitlength,page=3]{Figures/misalignments/misalignement-cartoons-y-rotated_v2.pdf}}%
  \end{picture}%
\endgroup%

        \hspace{10pt}
        \includegraphics[width = 0.74\linewidth]{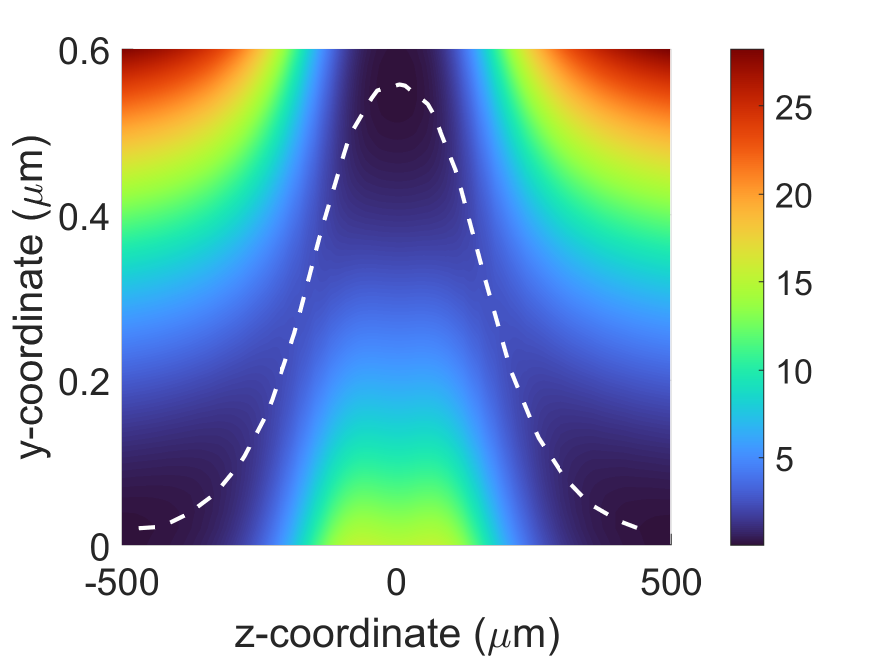}
        \label{fig:vert-offset-potmins}
    }
    \\
    \centering
    \subfloat[]{
        \def\svgwidth{0.2\linewidth}
\begingroup%
  \makeatletter%
  \providecommand\color[2][]{%
    \errmessage{(Inkscape) Color is used for the text in Inkscape, but the package 'color.sty' is not loaded}%
    \renewcommand\color[2][]{}%
  }%
  \providecommand\transparent[1]{%
    \errmessage{(Inkscape) Transparency is used (non-zero) for the text in Inkscape, but the package 'transparent.sty' is not loaded}%
    \renewcommand\transparent[1]{}%
  }%
  \providecommand\rotatebox[2]{#2}%
  \newcommand*\fsize{\dimexpr\f@size pt\relax}%
  \newcommand*\lineheight[1]{\fontsize{\fsize}{#1\fsize}\selectfont}%
  \ifx\svgwidth\undefined%
    \setlength{\unitlength}{84.37635707bp}%
    \ifx\svgscale\undefined%
      \relax%
    \else%
      \setlength{\unitlength}{\unitlength * \real{\svgscale}}%
    \fi%
  \else%
    \setlength{\unitlength}{\svgwidth}%
  \fi%
  \global\let\svgwidth\undefined%
  \global\let\svgscale\undefined%
  \makeatother%
  \begin{picture}(1,2.15584495)%
    \lineheight{1}%
    \setlength\tabcolsep{0pt}%
    \put(0,0){\includegraphics[width=\unitlength,page=1]{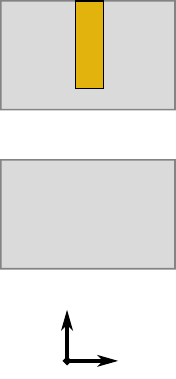}}%
    \put(0.27621638,0.37089504){\makebox(0,0)[lt]{\lineheight{1.25}\smash{\begin{tabular}[t]{l}x\end{tabular}}}}%
    \put(0.63455523,0.00000001){\makebox(0,0)[lt]{\lineheight{1.25}\smash{\begin{tabular}[t]{l}z\end{tabular}}}}%
    \put(0,0){\includegraphics[width=\unitlength,page=2]{Figures/misalignments/misalignement-cartoons-x.pdf}}%
    \put(0.04,0.8887391){\makebox(0,0)[lt]{\lineheight{1.25}\smash{\begin{tabular}[t]{l}5$\mum$\end{tabular}}}}%
  \end{picture}%
\endgroup%

        \hspace{10pt}
        \includegraphics[width = 0.74\linewidth]{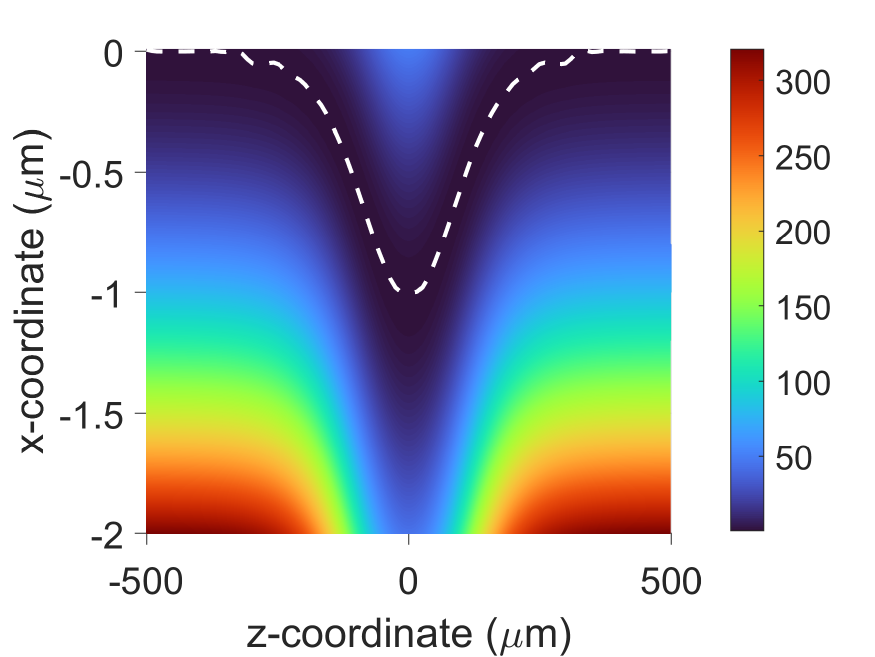}
        \label{fig:x-offset-potmins}
    }    
    \caption{The pseudopotential disfiguration in $\mu$eV when one of the FFPC side is translated by 5 $\mum$ along the z-direction (a), along y (b) and along x (c). The superimposed dashed white lines follow the pseudopotential minimum line. The sketches on the left show schematics of the misaligned FFPC for the corresponding 2D pseudopotential on the right. The grey blocks represent the rf electrodes whilst the gold blocks represent the FFPC.}
    \label{fig:misalignement}
\end{figure}

We introduce deliberate misalignment of the FFPC fibres and their shields to emulate experimental imperfection. Sources of errors include the production of the fibre facets as well as off-concentric positioning of the FFPC fibres in the shields. This inevitably breaks the aforementioned geometrical symmetry. First we look at the case of a relative offset of the FFPC sides by 5~$\mum$, which is typical experimental imperfection, along the $z$ axis. We find that this converts the straight potential minimum line to a curved one on the $xz$ plane as depicted in \refig{fig:ax-offset-potmin}. Albeit curved, the rf-null line is retained, i.e. there are no potential barriers as a result of this misalignment.

In the second case, we simulate the pseudopotential for a relative misalignment of the FFPC sides in the $y$ direction. Likewise, we find that the pseudopotential minimum line is skewed from the central axis in the vicinity of the trapping region of interest (see \refig{fig:vert-offset-potmins}). When one of the FFPC sides is raised up by 5~$\mum$, the pseudopotential minimum position at $z=0$ is also shifted up by about 550~nm. The ion at $z=0$ would also experience a relatively smaller shift of 20 nm in the $x$ direction. This is not shown in the plots for brevity.  Along the potential minimum line, a shallow potential is formed with barriers below 2.6 $\mu$eV at $z = \pm$ 174 $\mum$, with a secular frequency contribution of 3.1 kHz. This remains negligible compared to the axial dc confinement. Nonetheless, a Ca$^+$ ion positioned at the peak of this potential barrier would incur excess micromotion of about 40 nm. c.f. At a secular frequency of 400 kHz, the expected position spread for a Doppler cooled ion is $\sim$11 nm.

Lastly, we consider the case where there is a misalignment in the $x$ direction. One of the fibres is translated toward the trap center by 5~$\mum$. As can be seen in \refig{fig:x-offset-potmins}, the pseudopotential minimum line gets pulled toward the shifted fibre. The ions are resultantly trapped $\sim$1~$\mum$ away from the trap center. Here, as in the case of misalignment in the $z$ direction, the rf-null line in retained albeit curved. 

In conclusion, small geometrical symmetry breaking introduced by misalignments in the FFPC lead to minor deformation of the trap potential. In two of the cases (misalignment in the $x$ and $z$ directions), the deformation only comes in the form of a skewed potential minimum line, and does not result in potential bumps. The deformations are small enough in magnitude that they do not prohibit the successful overlapping of an ion with the cavity mode. In the other case (misalignment in $y$), the rf-null line is interrupted by small potential bumps which would subsequently lead to excess micromotion. Therefore misalignment in this direction should be disfavoured over the other two.  Due to experimental imperfection, optimising the cavity signal may require the introduction of a relative offset between the FFPC sides. However, the offset direction can be chosen by rotating one of the FFPC sides such that potential bumps are minimised.

\section{Cancellation of the rf field and trap symmetry}
\label{sec:trap-symmetry}

\subsection{General conditions}
\label{sec:trap-symmetry-gen-cond}

In Section \ref{sec:axial_potential}, we observed that the rf-field is no longer cancelled on the trap axis when the shielded FFPC is incorporated with the single-rf drive.
In this section, we analytically study the cancellation of the rf-field on the trap axis of a linear ion trap in general. We derive sufficient conditions for the symmetry of the electrodes placements and rf-drive type so that the cancellation of the rf-field on the trap axis is achieved in the absence of the translational symmetry.

First we consider a general case where we do not assume any particular geometry of the electrodes but only assume the electric field is produced by a charge distribution $\rho(x, y, z)$ where the trap axis coincides with the z-axis. The electric field at $z = z_0$ on the trap axis is 
\begin{equation}
\begin{split}
&\vec{E}(0, 0, z_0) =\\
&- \fr{1}{4\pi\epsilon_0}\grad\intd{x'}\!\!\intd{y'}\!\!\intd{z'}\\
&\eval{\fr{\rho(x', y', z')}{((x-x')^2+(y-y')^2+(z-z')^2)^{1/2}}}_{x=y=0,z=z_0}.
\end{split}
\end{equation}
Here $\grad$ is taken with respect to $x$, $y$ and $z$. Each component of $\vec{E}(0, 0, z_0)$ reads as follows.
\begin{align}
    E_x(z_0) &= \fr{-1}{4\pi\epsilon_0}\iiint  \fr{\rho(x', y', z')x'~dx'dy'dz'}{(x'^2+y'^2+(z'-z_0)^2)^{3/2}}, \label{eq:Ex_int} \\
    E_y(z_0) &= \fr{-1}{4\pi\epsilon_0}\iiint \fr{\rho(x', y', z')y'~dx'dy'dz'}{(x'^2+y'^2+(z'-z_0)^2)^{3/2}}, \label{eq:Ey_int} \\
    E_z(z_0) &= \fr{1}{4\pi\epsilon_0}\iiint \fr{\rho(x', y', z')(z'-z_0)~dx'dy'dz'}{(x'^2+y'^2+(z'-z_0)^2)^{3/2}} \label{eq:Ez_int},      
\end{align}
where we abbreviated $E_i(0, 0, z_0)\,(i = x, y, z)$ as $E_i(z_0)$ since we are only interested in the electric field on the trap axis. The conditions for vanishing E-field on the trap axis are $E_x(z_0) = E_y(z_0) = E_z(z_0)$ = 0:
\begin{align}
    \intd{x}\!\!\intd{y}\!\!\intd{z} \fr{\rho(x, y, z)x}{(x^2+y^2+(z-z_0)^2)^{3/2}} &= 0, \label{eq:cond-x} \\
    \intd{x}\!\!\intd{y}\!\!\intd{z} \fr{\rho(x, y, z)y}{(x^2+y^2+(z-z_0)^2)^{3/2}} & = 0, \label{eq:cond-y}\\
    \intd{x}\!\!\intd{y}\!\!\intd{z} \fr{\rho(x, y, z)(z-z_0)}{(x^2+y^2+(z-z_0)^2)^{3/2}} &= 0. \label{eq:cond-z} 
\end{align}
Regarding \refeq{eq:cond-x} and \refeq{eq:cond-y}, if the charge distribution has either the mirror symmetry around $x$ and $y$, \textit{i.e.} $\rho(x, y ,z) = \rho(-x, y ,z)$ and $\rho(x, y ,z) = \rho(x, -y ,z)$ or the parity symmetry $\rho(x, y ,z) = \rho(-x, -y ,z)$ in the $xy$ plane, the equations are satisfied. Regarding \refeq{eq:cond-z},  if the charge distribution satisfies the mirror symmetry around $z = z_0$, that is
\begin{align}
    \forall z, \  \rho(x, y, z_0+z) = \rho(x, y, z_0-z), \label{z-mirror}
\end{align}
then $\refeq{eq:cond-z}$ is automatically satisfied. Note that in order to have rf-null all along the trap axis, $z_0$ needs to be arbitrary in $\refeq{z-mirror}$. This actually means the translational symmetry $\forall z, z',\ \rho(x, y, z) = \rho(x, y, z')$. In our model, the electrode geometry holds the parity symmetry in the $xy$ and in the absence of the shielded FFPC it approximately satisfies the translational symmetry along $z$ apart from around the edges of the rf electrodes. Therefore in that case \refeq{eq:cond-x} and \refeq{eq:cond-y} are satisfied exactly and \refeq{eq:cond-z} is satisfied approximately to a good degree as was also demonstrated numerically in \refig{axial_pot_2D_no_shield} and \refig{axial_pot_1D_no_shield}.  

When a disruptive element such as a shielded FFPC is introduced, however, the translational symmetry is broken. Therefore \refeq{eq:cond-z} can no longer be satisfied based on the symmetry along the $z$-axis. This is the reason that the miscancelled rf-field only exhibit the $E_z$ components as in \refig{fig:Efield-components}. Now we consider a way to satisfy $\refeq{eq:cond-x}$-$\refeq{eq:cond-z}$ without relying on the translational symmetry along $z$. We consider the case where $\refeq{eq:cond-x}$-$\refeq{eq:cond-z}$ are already satisfied when the integrals over $x$ and $y$ are carried out before over $z$. That is, 
\begin{align}
    \intd{x}\!\!\intd{y} \fr{\rho(x, y, z)x}{(x^2+y^2+(z-z_0)^2)^{3/2}} &= 0, \label{eq:cond2-x}\\
    \intd{x}\!\!\intd{y} \fr{\rho(x, y, z)y}{(x^2+y^2+(z-z_0)^2)^{3/2}} &= 0, \label{eq:cond2-y}\\
    \intd{x}\!\!\intd{y} \fr{\rho(x, y, z)}{(x^2+y^2+(z-z_0)^2)^{3/2}} &= 0. \label{eq:cond2-z}
\end{align}
$\refeq{eq:cond2-x}$-$\refeq{eq:cond2-z}$ are sufficient conditions for $\refeq{eq:cond-x}$-$\refeq{eq:cond-z}$. Physically $\refeq{eq:cond2-x}$-$\refeq{eq:cond2-z}$ correspond to the situation where the E-field contribution from the charges on a single plane perpendicular to the trap axis vanishes independently from the contributions from other parallel planes (see \refig{fig:rod-trap-3d}). Let us rewrite $\refeq{eq:cond2-x}$-$\refeq{eq:cond2-z}$ in the polar coordinates $(r, \theta, z)$ with relationships $x = r\cos{\theta}$ and $y = r\sin{\theta}$: 
\begin{align}
    \intd{r}\!\!\intd{\theta} \fr{\rho(r, \theta, z)r\cos\theta}{(r^2+(z-z_0)^2)^{3/2}} &= 0, \label{cond2-x-polar}\\
    \intd{r}\!\!\intd{\theta} \fr{\rho(r, \theta, z)r\sin\theta}{(r^2+(z-z_0)^2)^{3/2}} &= 0, \label{cond2-y-polar}\\
    \intd{r}\!\!\intd{\theta} \fr{\rho(r, \theta, z)}{(r^2+(z-z_0)^2)^{3/2}} &= 0. \label{cond2-z-polar}
\end{align}
These conditions have to be met irrespective of $z$ and $z_0$. Since $z$ and $z_0$ only appear in conjunction with the integrals with respect to $r$, this leads to the following equations for the angular distribution of $\rho(r, \theta, z)$:
\begin{align}
    \intd{\theta} \rho(r, \theta, z)\cos\theta &= 0, \label{eq:cond3-x-polar}\\
    \intd{\theta} \rho(r, \theta, z)\sin\theta &= 0, \label{eq:cond3-y-polar}\\
    \intd{\theta} \rho(r, \theta, z) &= 0. \label{eq:cond3-z-polar}
\end{align}

\subsection{Case studies: 3D and planar traps}
\label{sec:trap-symmetry-cases}

\subsubsection{3D trap}
\begin{figure}[htb]
    \centering
    \subfloat[]{\includegraphics[width=0.8\linewidth]{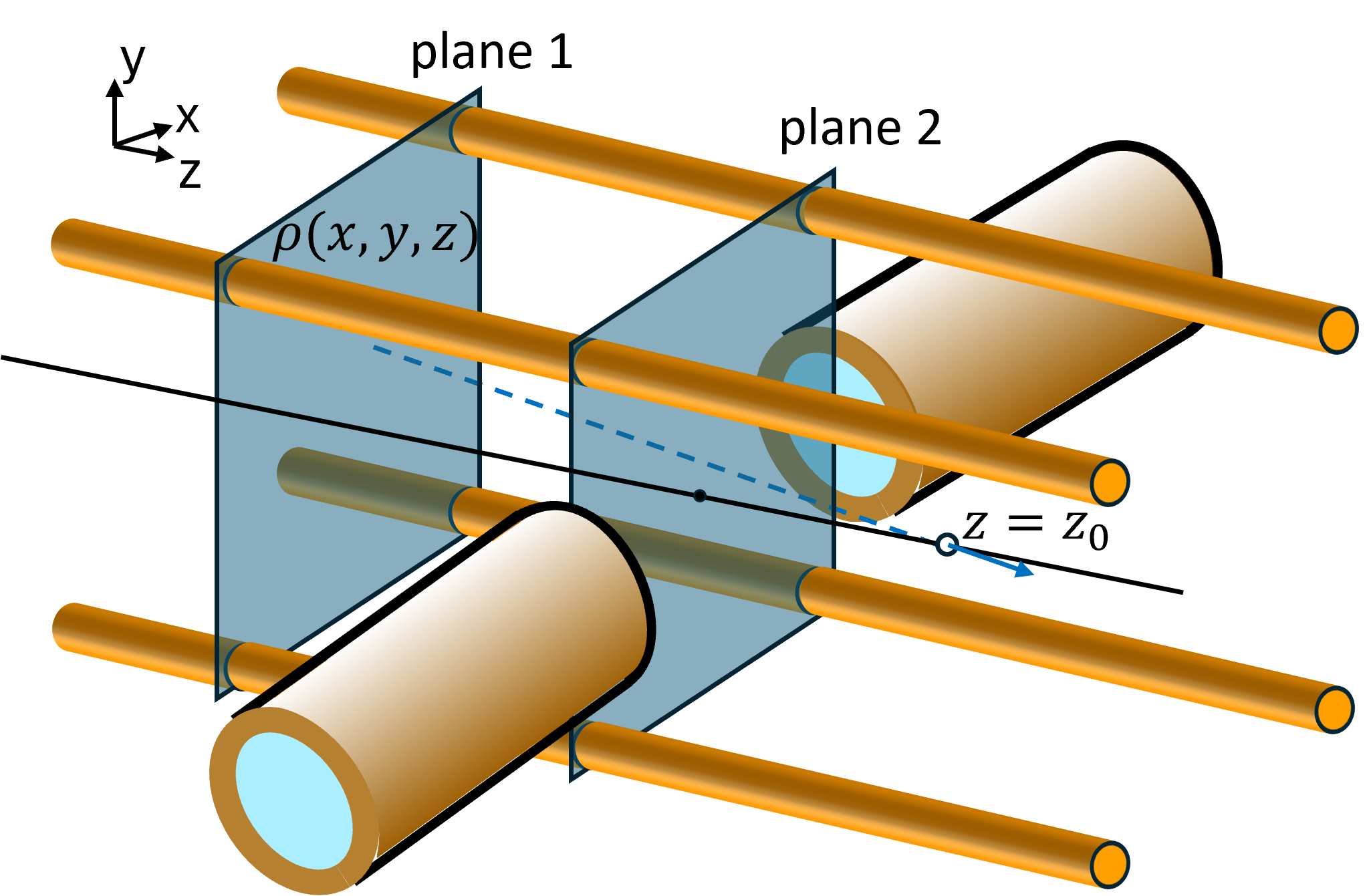}
        \label{fig:rod-trap-3d}}\\
    \subfloat[]{\includegraphics[width=0.8\linewidth]{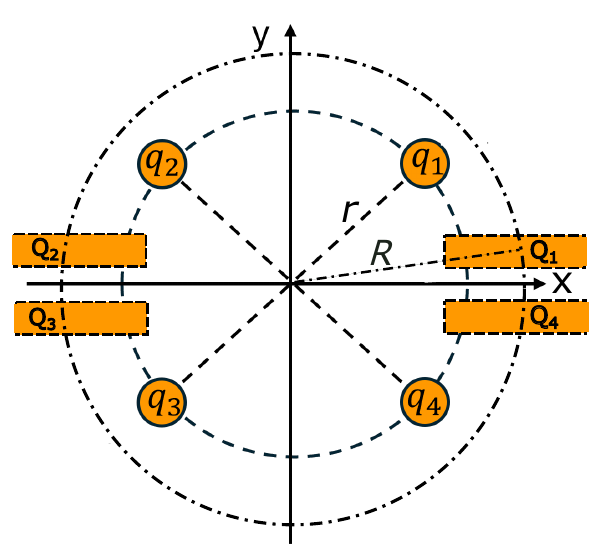}
        \label{fig:rod-trap-xy}}
    \caption{\textbf{(a)} Linear trap with four rods electrodes and two fibre shields. \textbf{(b)} Cross section of a plane perpendicular to the trap axis. The cross sections of the rods are at a distance to the centre of $r$ possess charges $q_1$, $q_2$, $q_3$ and $q_4$ respectively. The cross sections of the shields at a distance to the centre of $R$ possess charges $Q_1$, $Q_2$, $Q_3$ and $Q_4$ respectively.} 
    \label{fig:rod-trap} 
\end{figure}

As a three-dimensional linear trap, for the brevity of the analysis we consider a traditional four-rod configuration as shown in \refig{fig:rod-trap-3d}. The four identical rods are located at the equal distance from the trap axis along $z$. Note that our blade-style trap can be approximated to this form by regarding the tips of the blades work similar to the rods. Fibres with shields are placed symmetrically along one of the transverse directions $x$.

Consider the cross section of the conductive materials with a plane perpendicular to the trap axis. The charge distribution on any such plane should satisfy \refeq{eq:cond3-x-polar}-\refeq{eq:cond3-z-polar}. First we consider a plane that does not intersect the fibre shields such as the plane 1 in \refig{fig:rod-trap-3d}. There are four points of the rods intersecting the plane where they are approximated as point charges $q_1$, $q_2$, $q_3$ and $q_4$. The distance from the rod intersections to the trap axis is $r$. Then $\refeq{eq:cond3-x-polar}$, $\refeq{eq:cond3-y-polar}$ and $\refeq{eq:cond3-z-polar}$ are rewritten in terms of  $q_1$, $q_2$, $q_3$ and $q_4$ as follows:
\begin{align}
    q_1 -q_2-q_3+q_4 = 0, \label{eq:rod-cond-x}\\
    q_1 +q_2-q_3-q_4 = 0, \label{eq:rod-cond-y}\\
    q_1 +q_2+q_3+q_4 = 0, \label{eq:rod-cond-z}
\end{align}
which lead to
\begin{align}
    q_1 = q_3 = -q_2 = -q_4. \label{eq:rod-cond}
\end{align}

Next we investigate the electric potential that fulfils the charge condition \refeq{eq:rod-cond} as it is the electric potential not the charge distribution that one can directly control in the lab. In general, the charge can be linearly related to the potential distribution as follows:
\begin{align}
    q_i &= \int\!\phi(r',\theta',z')g_i(z; r', \theta', z') \,dr'd\theta' dz' 
    \quad (i=1,2,3,4), \label{eq:charge-greens}
\end{align}
where $z$ is the coordinate of the plane of interest. $g_i(z; r', \theta', z')$ is similar to Green's function for the Poisson equation $\nabla^2\phi = -\rho/\epsilon$. However here we are solving the inverse problem: the charge is derived from the potential. Physically $g_i(z; r', \theta', z')$ corresponds to the charge on the rod $i$ at $z$ when the potential is set to be a delta function located at $(r',\theta', z')$.
One can easily find the following relations between the four functions, which are true when the fibres and the shields are placed symmetrically along the $x$ direction as in \refig{fig:rod-trap-3d}:
\begin{align}
&g_1(z; r', \theta', z') = g_2(z; r', \pi-\theta', z') \nonumber\\
&\quad = g_3(z; r',\theta'-\pi, z') = g_4(z; r', -\theta', z')
\label{eq:Greens-relation}
\end{align}
Substituting the equation \refeq{eq:charge-greens} in \refeq{eq:rod-cond} and using the relations \refeq{eq:Greens-relation}, one obtains
\begin{align}
&\int\!\phi(r', \theta', z')g_1(z; r', \theta', z') \,dr'd\theta' dz' \nonumber\\
&= -\int\!\phi(r', \pi-\theta', z')g_1(z; r', \theta', z') \,dr'd\theta' dz' \nonumber\\
&= \int\!\phi(r', \theta'-\pi, z')g_1(z; r', \theta', z') \,dr'd\theta' dz' \nonumber \\
&=  -\int\!\phi(r', -\theta', z')g_1(z; r', \theta', z') \,dr'd\theta' dz'.
\end{align}
These equations must hold irrespective of the dependence of $g_1(z; r', \theta', z') $ on $z$. Therefore they lead to the following conditions for the potential:
\begin{align}
    &\phi(r, \theta, z) = -\phi(r, \pi-\theta, z) \nonumber\\
    &= \phi(r, \theta-\pi, z) = -\phi(r, -\theta, z). \label{eq:pot-cond-rod}
\end{align}
It is clear that the single-rf drive does not meet these conditions but the dual-rf drive plus grounded shields does.

Next we consider a plane that intersects the fibre shields as well as the rods (the plane 2 in \refig{fig:rod-trap-3d}). The intersection of the shields with the plane is shown as the rectangular regions in \refig{fig:rod-trap-xy}. We consider a circle with a radius $R$ on this plane as depicted in \refig{fig:rod-trap-xy}. We denote the charges at the intersections of the shields and the circle as $Q_1$, $Q_2$, $Q_3$ and $Q_4$. The charges on this circle must satisfy \refeq{eq:cond3-x-polar}-\refeq{eq:cond3-z-polar} irrespective of $R$.
When $R\neq r$, the circle does not intersect the rods and hence the conditions for the charge distribution are given by the same form as \refeq{eq:rod-cond-x}-\refeq{eq:rod-cond-x} with $q_i$ replaced by $Q_i$. Then we obtain
\begin{equation}
Q_1 = Q_3 = -Q_2 = -Q_4,
\end{equation}
which is essentially the same requirement as \refeq{eq:rod-cond} and leads to \refeq{eq:pot-cond-rod}.

When $R = r$, now the circle intersects both rods and shields. $\refeq{eq:cond3-x-polar}$-$\refeq{eq:cond3-z-polar}$ are rewritten as 
\begin{align}
    &q_1 -q_2-q_3+q_4 + \gamma(Q_1 -Q_2-Q_3+Q_4) = 0, \label{eq:Q-and-q-x}\\
    &q_1 +q_2-q_3-q_4 + \delta(Q_1 +Q_2-Q_3-Q_4)  = 0, \label{eq:Q-and-q-y}\\
    &q_1 +q_2+q_3+q_4 + Q_1 +Q_2+Q_3+Q_4 = 0, \label{eq:Q-and-q-z}
\end{align}
where $\gamma = \frac{\cos\beta}{\cos\alpha}$ and $\delta = \frac{\sin\beta}{\sin\alpha}$ with $\alpha$ being the angle between the charge $q_1$ and the $x$ axis, $\beta$ the angle between $Q_1$ and the $x$ axis. An obvious solution for these conditions is such that $q_i$ and $Q_i\,(i = 1, 2, 3, 4)$ satisfy the following equations individually. 
\begin{align}
    q_1 = q_3 = -q_2 = -q_4, \label{eq:rod-cond-q}\\
    Q_1 = Q_3 = -Q_2 = -Q_4.  \label{eq:rod-cond-Q}
\end{align}
Then the same line of argument as above can be applied to both $q_i$ and $Q_i\,(i = 1, 2, 3, 4)$ in order to obtain necessary electric potential. Then one is led to the same conclusion \refeq{eq:pot-cond-rod} which can be met by the dual-rf drive but not by the single-rf drive. 

In conclusion, we have shown that in a 3D linear ion trap with grounded shields and the dual-rf drive, charges on a individual plane, whether it is outside or inside the cavity region, always result in the cancellation of the electric field on the trap axis. However this is not the case with the single-rf drive.     

\subsubsection{Surface trap}
\begin{figure}[htb]
    \centering
    \subfloat[]{\includegraphics[width=0.8\linewidth]{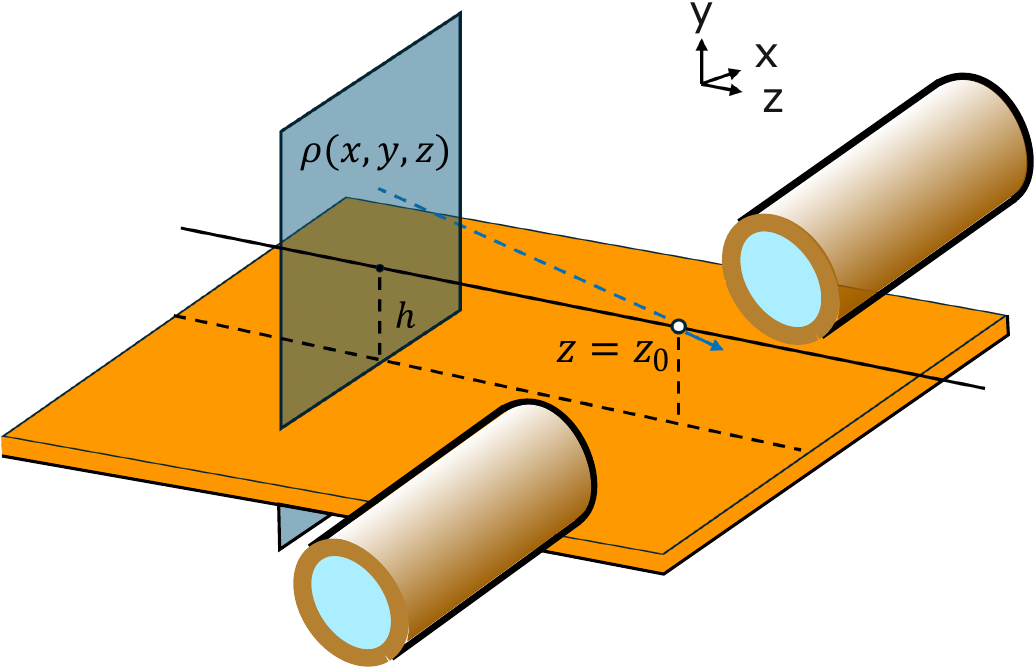}
        \label{fig:surface-trap-3d}}\\
    \subfloat[]{\includegraphics[width=0.8\linewidth]{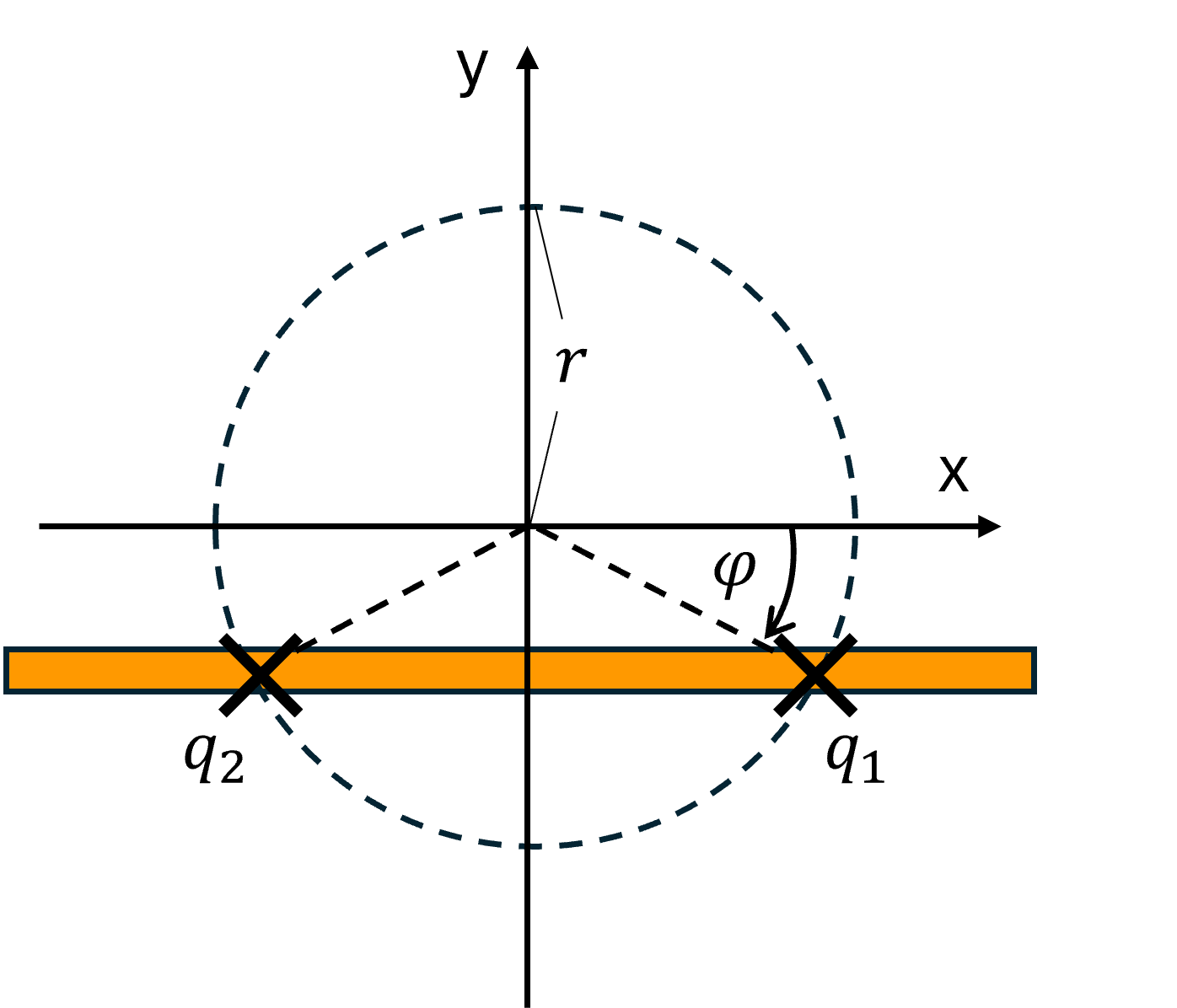}
        \label{fig:surface-trap-xy}}
    \caption{\textbf{(a)} Planar trap modeled as a plane with an ion height $h$. \textbf{(b)} Cross section of the planar trap with a plane perpendicular to the trap axis. Intersection of the surface trap plane and a circle in the plane of reference is depicted by the crosses.} 
    \label{fig:surface-trap} 
\end{figure}

Next we consider a geometry that generalizes the planar surface linear trap. We will show that no charge distribution configuration can creates linear rf-null. Here we do not assume a particular electrode configuration but only require that the charges on the surface trap are confined in a plane parallel to the $xz$ plane (\refig{fig:surface-trap-3d}). The plane is located at a distance $h$ from the trap axis corresponding to the height of the ion. We consider the cross-section of a plane perpendicular to the $z$ axis that does not intersect with the fibre shield (see \refig{fig:surface-trap-3d}). Since equations $\refeq{eq:cond3-x-polar}$, $\refeq{eq:cond3-y-polar}$ and $\refeq{eq:cond3-z-polar}$ should hold on any plane perpendicular to the trap axis, choosing such a plane simplifies the proof. For a given radius of a circle $r$, there are two points of intersection at an angle $-\varphi$ and $\varphi-\pi$ respectively with $\varphi = \asin(h/r)$ (see \refig{fig:surface-trap-xy}). We again approximate them as point charges $q_1$ and $q_2$ respectively. 
$\refeq{eq:cond3-x-polar}$, $\refeq{eq:cond3-y-polar}$ and $\refeq{eq:cond3-z-polar}$ are rewritten with $q_1$, $q_2$ and $\varphi$ as follows:
\begin{align}
    q_1\cos\varphi  - q_2\cos\varphi = 0, \\
    q_1\sin\varphi +q_2\sin\varphi = 0, \\
    q_1 +q_2 = 0.
\end{align}
These lead to
\begin{align}
    q_1-q_2 = 0, \\
    q_1+q_2 = 0,
\end{align}
and then a trivial solution of $q_1 = q_2 = 0$ follows. Therefore in the case of the surface trap, there is no charge distribution that creates a trapping potential whilst satisfying $\refeq{eq:cond3-x-polar}$, $\refeq{eq:cond3-y-polar}$ and $\refeq{eq:cond3-z-polar}$ simultaneously.
\begin{figure}[h!]
    \centering
    \includegraphics[width=1\linewidth]{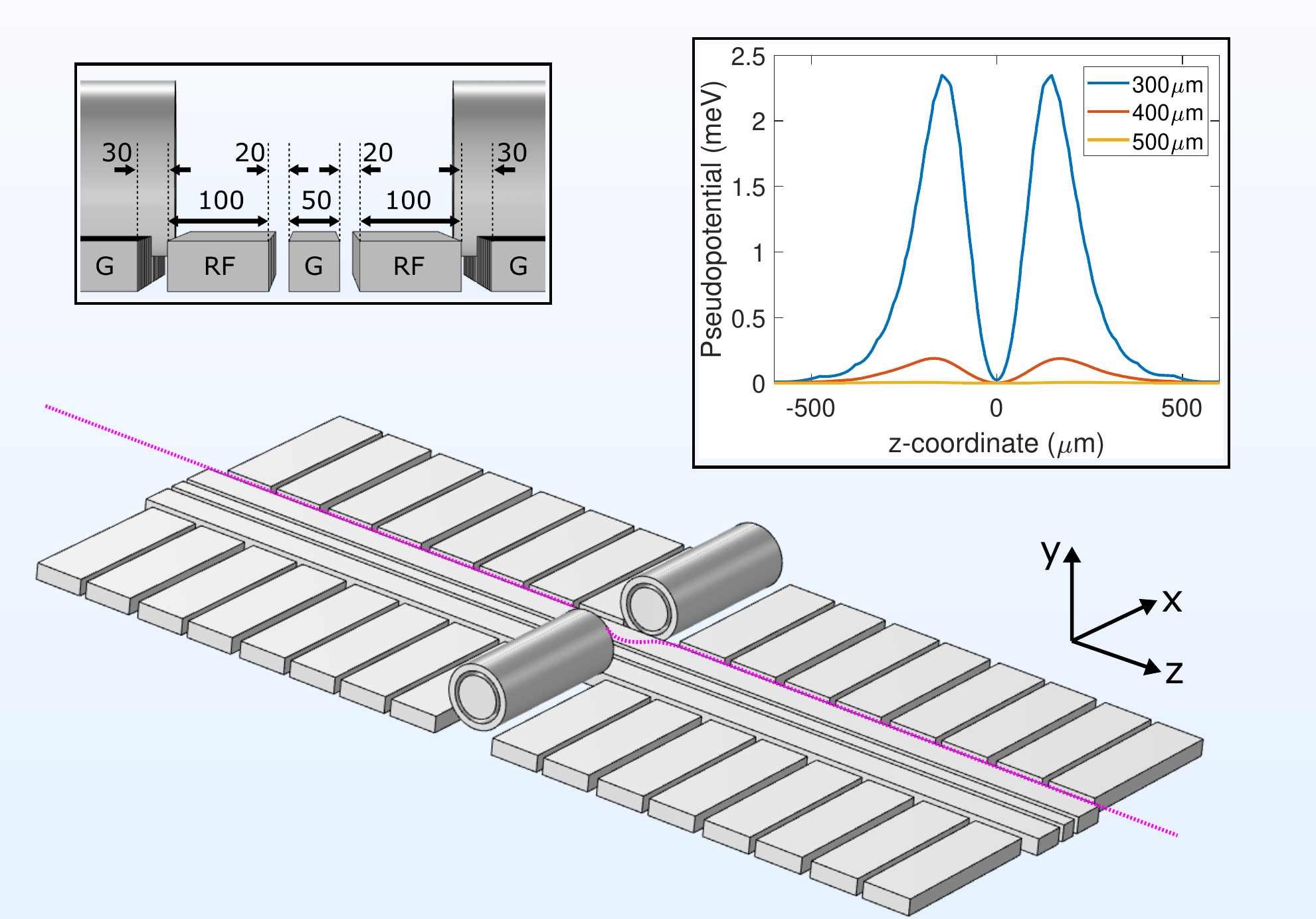}
    \caption{The modelled surface ion trap. The purple dashed line is the trapping trajectory on the potential minimum (the centre distortion exaggerated). The left inset shows the surface trap viewed from the $xy$ plane with relevant dimensions in $\mum$. The label `G' stands for ground. The pseudopotential shown in the right inset is along the trapping trajectory with various cavity lengths.}
    \label{fig:surf-compare}
\end{figure}

\begin{figure}[htb]
    \centering
    \includegraphics[width=0.8\linewidth]{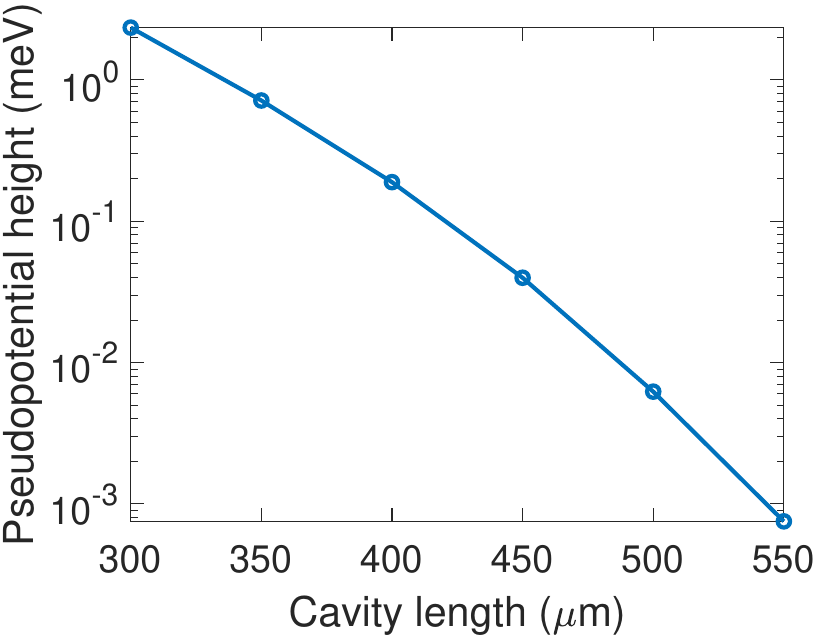}
    \caption{The height of the pseudopotential barrier near the FFPC as a function of the cavity length.}
    \label{fig:surf-height-cavlen}
\end{figure}

It should be pointed out that unlike 3D trap, the ion chain is not necessarily a straight line after introducing the fibre even without misalignment (we will see this later in the simulation). However, it is easy to prove that the rf cannot be cancelled along the deformed pseudopotential minimum either, because the proof above is valid for any ion height $h$. 

Similar to the 3D trap with the single rf drive, violating the electric field cancellation condition distorts the pseudo-potential in a surface trap. It is important to know if such distortion is experimentally significant. We simulate a surface trap depicted in \refig{fig:surf-compare} (see \cite{David2023} for a similar design). Two rf rail-electrodes on both sides of a ground rail with an applied voltage of 25~V are used for generating a linear trapping potential (see the left inset in \refig{fig:surf-compare} for the electrode dimensions). The shielded FFPC has the same dimensions as in the previous simulations. The trap depth without an FFPC is $\sim$27~meV, a typical trap depth for a surface trap. When an FFPC is present, the potential minimum line is distorted slightly, shown as the purple dashed line in \refig{fig:surf-compare}. Along this line, a pseudopotential well is found near the FFPC as in the case II of Section~\ref{sec:price}, with the depth of the well being $\sim$2.3~meV when the cavity length is 300~$\mum$ (see the right inset in \refig{fig:surf-compare}). In this potential well only a single ion can be trapped without excess micromotion. For $^{40}\text{Ca}^+$, 2.3~meV corresponds to an ion speed $v=$~105~m/s of the kinetic energy $mv^2/2$ and a motional temperature of 27~K. Such potential well prevents ions from shuttling in and out of the cavity region. The depth is reduced to 189 $\mu$eV when the FFPC and the shield are retracted to a cavity length of 400~$\mum$. and 6.2~$\mu$eV with 500~$\mum$ cavity length (see \refig{fig:surf-height-cavlen}).  Ion shuttling against such low potential barriers becomes possible. However 500~$\mum$ of cavity length also makes the coupling of the ion to the cavity smaller and the cavity alignment more challenging.

Compared with 3D traps, in surface traps the rf-null along the trap axis is inevitably compromised to a single-spot rf null due to the lack of geometric symmetry. Nonetheless, thanks to the relatively shallow trapping depths, the pseudopotential barriers along the potential minimum line can be made low enough for ion shuttling to be possible if the FFPC is retracted. However the retraction comes with drawbacks as mentioned above. The simulation shows that such trade-off is at the edge of practicality. Therefore, an ion-cavity experiment with a surface trap should be designed carefully considering both the pseudopotential walls and the drawbacks with the cavity retraction.

\section{Discussion and conclusion}
\label{sec:discussion}

In this paper, we have studied the integration of an FFPC in a linear ion trap with a focus on the shielding of the fibre dielectrics. We have found out that employing conductive shields around the FFPC is advantageous in mitigating the effects of stray charges and reducing the motional heating. For the latter, retracting the fibres inside the shield shows enhanced distance dependence for the heating rate in comparison with the case without a shield. 
However, in the case of the conventional single-rf drive, such conductive shields induce miscancellation of the rf field and introduce undesired pseudopotential bumps along the trap axis. One simple solution is to use the dual-rf drive, which guarantees the linear rf-null in a 3D trap. On the other hand, such a driving scheme cannot be implemented on the single-sided trap geometry such as in a surface trap, which makes the integration of an FFPC with a surface trap technically more challenging

From these observations, it can be conjectured that the following elements would be essential for successful integration of an FFPC in a linear ion trap: 1) Conductive shielding of the fibre dielectrics, 2) use of a miniature 3D trap and the dual-rf drive to avoid the rf-miscancellation and 3) retraction of the fibres inside the shields to reduce the heating rate. Regarding miniature 3D traps, miniaturization is required because the cavity mirrors need to be placed outside the trapping region. In contrast to planar surface traps where conventional micro-lithgraphy is available, it used to be technically difficult to construct a 3D electrode structure with micrometer accuracy. Recently, however, several 3D microfabrication technologies have become available and they have been applied to ion traps successfully \cite{ragg2019segmented,xu20233d,wang2020coherently}. By using these technologies, a 3D ion trap suitable for FFPC integration can be fabricated \cite{Soon2024}. Regarding the third element in the above list, the heating from dielectirc fibre surfaces would be a substantial obstacle for scalable ion-based quantum processors even if stable trapping is achieved in a FFPC. Our finding in Section~\ref{subsec:heating} shows that one can modify the distance scaling of the heating rate by simply retracting the fibres inside the shields. Furthermore ion shuttling can be used to bring the net heating rate even lower. With these techniques, it is anticipated that a miniature optical cavity and motion-based quantum gate operations can co-exist in the same ion trap without resorting to a cryogenic environment.

While 3D ion traps are an ideal platform for the miniature cavity integration, if one opts for surface traps for their greater manufacturing flexibility, one avenue that can be derived from our analysis is to use a relatively long cavity length on the order of $\sim$500~$\mum$ to avoid pseudopotential bumps caused by the rf miscancellation. However the expected heating rate from the fibre dielectrics becomes higher than in the 3D trap at the same cavity length since whilst in the latter the cavity length can be achieved by retracting the fibres in the shields, in the surface trap the shields have to be retracted together as they are the sources of the pseudopotential bumps. One can suppress the heating rate by extensive use of ion shuttling in and out of the cavity region or by operating the system at a cryogenic temperature.  

In conclusion, this paper has thoroughly analysed the problem of the integration of a miniature cavity in a linear ion trap in several novel perspectives. The integration of a miniature cavity in a linear ion trap is a paramount milestone in the trapped-ion based quantum technologies but its realization has been eluded for decades despite some experimental efforts. We regard that one of the difficulties in this challenge has been a lack of clarity as to what are the essential obstacles and solutions therefor. As discussed above, this paper has pointed them out and proposed promising guidelines based on concrete numerical and analytical evidences. Future work should follow these guidelines, implement the integration of a FFPC in a linear ion trap, and examine the results presented in this paper experimentally.

\section{Acknowledgements}

This work was supported by JST Moonshot R\&D (Grant Number JPMJMS2063) and MEXT Quantum Leap Flagship Program (MEXT Q-LEAP)
(Grant Number JPMXS0118067477).
Simulations carried out using COMSOL MultiPhysics. We acknowledge funding from OIST Proof of Concept Program - Seed Phase Project (R$11\_59$).


\bibliography{LinTrapnCav}

\end{document}